\documentclass[11pt,a4paper,english,amssymb,nofootinbib,superscriptaddress]{revtex4-2}
\usepackage{lmodern}
\usepackage{lmodern}

\usepackage[T1]{fontenc}
\usepackage[latin9]{inputenc}
\usepackage{color}
\usepackage{babel}
\usepackage{calc}
\usepackage{amsmath}
\usepackage{amssymb}
\usepackage{graphicx}
\usepackage{esint}
\usepackage[pdfusetitle,
 bookmarks=true,bookmarksnumbered=false,bookmarksopen=false,
 breaklinks=false,pdfborder={0 0 1},backref=false,colorlinks=false]
 {hyperref}

\makeatletter

\providecommand{\tabularnewline}{\\}

\usepackage{babel}

\newcommand{\Riem}{R}

\newcommand{\cD}{\nabla}

\def\be{\begin{equation}}
\def\ee{\end{equation}}

\@ifundefined{textcolor}{}{%
	\definecolor{BLACK}{gray}{0}
	\definecolor{WHITE}{gray}{1}
	\definecolor{RED}{rgb}{1,0,0}
	\definecolor{GREEN}{rgb}{0,1,0}
	\definecolor{BLUE}{rgb}{0,0,1}
	\definecolor{CYAN}{cmyk}{1,0,0,0}
	\definecolor{MAGENTA}{cmyk}{0,1,0,0}
	\definecolor{YELLOW}{cmyk}{0,0,1,0}
}

\usepackage{latexsym}\usepackage{bm}

\makeatother

\begin{document}
\title{Geometry is Wavy: Curvature Wave Equations for Generic Affine Connections}
\author{Emel Altas}
\email{emel.altas@agu.edu.tr}

\affiliation{Department of Engineering Sciences,
 Abdullah Gul University, 38080 Kayseri, T\"{u}rkiye.}
\author{Bayram Tekin}
\email{bayram.tekin@bilkent.edu.tr}
\affiliation{Department of Physics,
 Bilkent University, 06800, Ankara,T\"{u}rkiye.}
\date{\today}
\begin{abstract}
\noindent Geometry is wavy: even at the purely geometric level (no
particular theory chosen), curvature satisfies a covariant quasilinear
wave equation. In Riemannian geometry equipped with the Levi-Civita
connection, the Riemann curvature tensor obeys a wave equation of
the schematic form 
\[
\Box Riem=\mathcal{Q}(Riem,Riem),
\]
where $\mathcal{Q}(Riem,Riem)$ denotes the terms quadratic in the
curvature arising from the Bianchi identities. In this work, we generalize
this curvature wave equation to spacetimes endowed with a generic
affine connection possessing torsion and nonmetricity. Working within the metric-affine framework, we derive the corresponding wave equation
for the Riemann tensor and analyze its structure in several geometrically
and physically distinguished settings, including Einstein spaces,
teleparallel gravity, and Einstein-Cartan theory. 
\end{abstract}
\maketitle
\vspace{0.8em}
 
\begin{center}
{\small\textit{Dedicated to Metin~G\"{u}rses on the occasion of his becoming
an emeritus professor after 60 years of research and teaching.}}\textit{ }
\par\end{center}

\vspace{0.8em}

\tableofcontents{}\clearpage{}

\section{INTRODUCTION}

In her beautiful and compact book \cite{ChoquetBruhat2015}, Choquet--Bruhat
shows that the Riemann curvature tensor, in the standard geometric setting of a torsion-free and metric-compatible connection, satisfies
a nonlinear second-order wave equation of the form 
\begin{align}
\cD^{\alpha}\cD_{\alpha}\Riem_{\beta\gamma\lambda\mu}+S_{\beta\gamma\lambda\mu}+\Big\{\cD_{\gamma}\big(\cD_{\lambda}\Riem_{\mu\beta}-\cD_{\mu}\Riem_{\lambda\beta}\big)-(\beta\to\gamma)\Big\}=0,\label{eq:I1412}
\end{align}
where the quadratic term in the curvature is given as 
\begin{align}
S_{\beta\gamma\lambda\mu} & =\Big\{\Riem_{\gamma}{}^{\rho}\,\Riem_{\rho\beta\lambda\mu}+\Riem^{\alpha}{}_{\gamma}{}^{\rho}{}_{\beta}\,\Riem_{\alpha\rho\lambda\mu}+\Big[\,\Riem^{\alpha}{}_{\gamma}{}^{\rho}{}_{\lambda}\,\Riem_{\alpha\beta\rho\mu}-(\lambda\to\mu)\Big]\Big\}-(\beta\to\gamma).\label{eq:I149}
\end{align}
This remarkable equation is purely geometric: it holds in dimensions
greater than three without invoking any field equations, and follows
solely from the differential Bianchi identities. In three dimensions, however, the differential Bianchi identity does not carry independent
content beyond the contracted one; consequently, the curvature-wave equation does not arise. The metric signature also plays no essential
role at this stage: in Riemannian signature, the same equation becomes
a nonlinear elliptic equation rather than a hyperbolic one.

An important feature of \eqref{eq:I1412} is that its trace does not
yield a nontrivial equation; in particular, the Ricci tensor does
not satisfy a closed wave equation on its own. Nevertheless, once the geometry is assumed to obey the Einstein field equations, 
\begin{equation}
G_{\alpha\beta}+\Lambda g_{\alpha\beta}=\kappa\,T_{\alpha\beta},\label{eq:EinsteinLambda}
\end{equation}
with $G_{\alpha\beta}:=R_{\alpha\beta}-\tfrac{1}{2}g_{\alpha\beta}R$
and $\kappa=8\pi G$ (in units $c=1$), the Ricci tensor becomes algebraically
determined by the energy-momentum tensor: 
\begin{equation}
R_{\alpha\beta}=\kappa\left(T_{\alpha\beta}-\frac{1}{2}g_{\alpha\beta}T\right)+\Lambda g_{\alpha\beta}.\label{eq:Ricci_in_terms_of_Ttilde}
\end{equation}
Introducing the trace-reversed energy-momentum tensor 
\begin{equation}
\widetilde{T}_{\alpha\beta}:=T_{\alpha\beta}-\frac{1}{2}g_{\alpha\beta}T,\label{eq:Ttilde_def}
\end{equation}
this relation takes the compact form 
\begin{equation}
R_{\alpha\beta}=\kappa\,\widetilde{T}_{\alpha\beta}+\Lambda g_{\alpha\beta}.\label{eq:Ricci_kappa_Ttilde_Lambda}
\end{equation}
Substituting this into \eqref{eq:I1412}, the curvature-wave equation
reduces to 
\begin{equation}
\nabla^{\alpha}\nabla_{\alpha}R_{\beta\gamma\lambda\mu}+S_{\beta\gamma\lambda\mu}+\kappa\Big\{\nabla_{\gamma}\big(\nabla_{\lambda}\widetilde{T}_{\mu\beta}-\nabla_{\mu}\widetilde{T}_{\lambda\beta}\big)-(\beta\to\gamma)\Big\}=0.\label{eq:wave_Riemann_EinsteinLambda}
\end{equation}
In a vacuum, the source terms in the curly brackets vanish identically; the first two terms survive and constitute a quasilinear wave equation
satisfied by the Riemann tensor.

While it is not entirely clear who first derived this curvature-wave
equation in the history of differential geometry, an early reference
is \cite{Ryan}, where it appears under the name of the \emph{Penrose wave equation} and is employed to provide an elegant 
derivation of the Teukolsky equation governing perturbations of the Kerr geometry.
More recently, a linearized version of this equation was used in \cite{Garfinkle}
to analyze the gravitational wave memory effect. Despite these appearances, we believe that the wave equation obeyed by the Riemann curvature
tensor deserves much wider attention, as it directly encodes the intrinsically
wavelike character of geometry in Lorentzian signature and its elliptic
character in Euclidean signature.

In this work, we extend this discussion beyond the Riemannian setting
by allowing for a general affine connection that is neither symmetric
nor metric-compatible. Although the resulting computations are considerably
more involved, the geometric and physical insights gained from the
general result justify the effort. Before presenting these results,
we briefly review the relevant geometric frameworks. Our conventions
are those of \cite{Wald}.

Einstein's general theory of relativity \cite{Einstein1915} describes
gravitation as a manifestation of spacetime curvature, encoded in
the Riemann tensor constructed from the Levi-Civita connection of
a Lorentzian metric. The differential geometry used in Einstein's
theory goes back to \cite{Riemann} and \cite{RicciLeviCivita}. This
formulation rests on two assumptions: the absence of torsion and the
metric compatibility of the affine connection. While natural and economical,
these assumptions are not imposed by first principles. From a differential-geometric
viewpoint, the most general affine connection on a manifold may possess
torsion and nonmetricity in addition to curvature.

Relaxing these assumptions leads to the framework of metric-affine
geometry, where the metric and affine connection are treated as independent
structures. In this setting, gravitational interactions may be encoded
not only in curvature but also in torsion and nonmetricity. Torsion
measures the antisymmetric part of the connection and governs the
failure of infinitesimal parallelograms to close, while nonmetricity
quantifies the variation of the metric under parallel transport. Together
with curvature, these tensors provide a complete local characterization
of general affine geometry.

Several important gravitational theories arise as consistent restrictions
of the metric-affine framework. General relativity is recovered when
both torsion and nonmetricity vanish. Einstein--Cartan theory \cite{Cartan}
allows for nonvanishing torsion while preserving metric compatibility,
thereby coupling spacetime geometry to the intrinsic spin of matter
\cite{HehlReview}. Teleparallel theories \cite{AldrovandiPereira,Adak},
by contrast, are characterized by vanishing curvature, with gravitation
encoded entirely in torsion and/or nonmetricity. In particular, metric
teleparallel gravity employs torsion alone \cite{AldrovandiPereira},
symmetric teleparallel gravity employs nonmetricity alone \cite{BeltranJimenez,Adak-Sert,M_Adak,Adak-Kalay},
and more general teleparallel models allow for both.

These interrelations are most transparently revealed through the decomposition
of a general affine connection into its Levi-Civita \cite{RicciLeviCivita}
part and tensorial corrections associated with torsion and nonmetricity.
This geometric landscape shows that gravity may be equivalently described
in terms of curvature, torsion, or nonmetricity, depending on the
chosen representation. See Figure 1 for a classification of these
theories.\footnote{We owe the last line of the table to F.W. Hehl, see the nice exposition \cite{Hehl0}}

Adopting the metric-affine perspective, we derive the wave equation
satisfied by the Riemann curvature tensor in the presence of both
torsion and nonmetricity. We then specialize our general results to
several geometrically and physically distinguished settings, including
Einstein spaces, teleparallel gravity \cite{AldrovandiPereira}, and
Einstein--Cartan theory \cite{Cartan}. This work highlights how
additional geometric structures modify the propagation and dynamics
of spacetime curvature, and it is dedicated to \textbf{Metin G\"{u}rses}
on the occasion of his becoming an emeritus professor in recognition
of his profound contributions to the geometry of gravitation.

\noindent\textbf{Outline of the paper:} In Section \ref{sec:preliminaries}
we fix conventions and summarize the metric-affine geometric ingredients
used throughout (torsion, nonmetricity, and the decomposition of a
generic connection into Levi-Civita plus distortion). Section~\ref{sec:curvature_decomposition}
derives the curvature, the Ricci tensor, and the scalar curvature
in terms of these fields and discusses which familiar Riemannian symmetries
survive in the generic case. In Section \ref{sec:identities} we establish
the generalized Ricci identities and Bianchi identities needed for
consistency checks and for manipulating covariant derivatives. Section
\ref{sec:dynamics} presents the field equations and the main results
of the paper, together with the limits that recover standard general
relativity. The appendices collect technical derivations and complementary
material: Appendix A introduces the decomposition of the generic affine
connection into the Levi Civita connection, torsion, and nonmetricity
contributions. Appendix B gives the decomposition of the curvature
tensors. Appendix C provides the commutator (Ricci) identities; Appendix
D derives the generalized Bianchi identities; Appendix E discusses
the teleparallel and symmetric-teleparallel specializations \cite{AldrovandiPereira},
while Appendix F summarizes the tetrad and spin connection \cite{HehlReview}
formulation and its relation to the metric-affine variables. The last appendix is devoted to a short historical survey.

\begin{figure}[t]
\centering \includegraphics[width=1\linewidth]{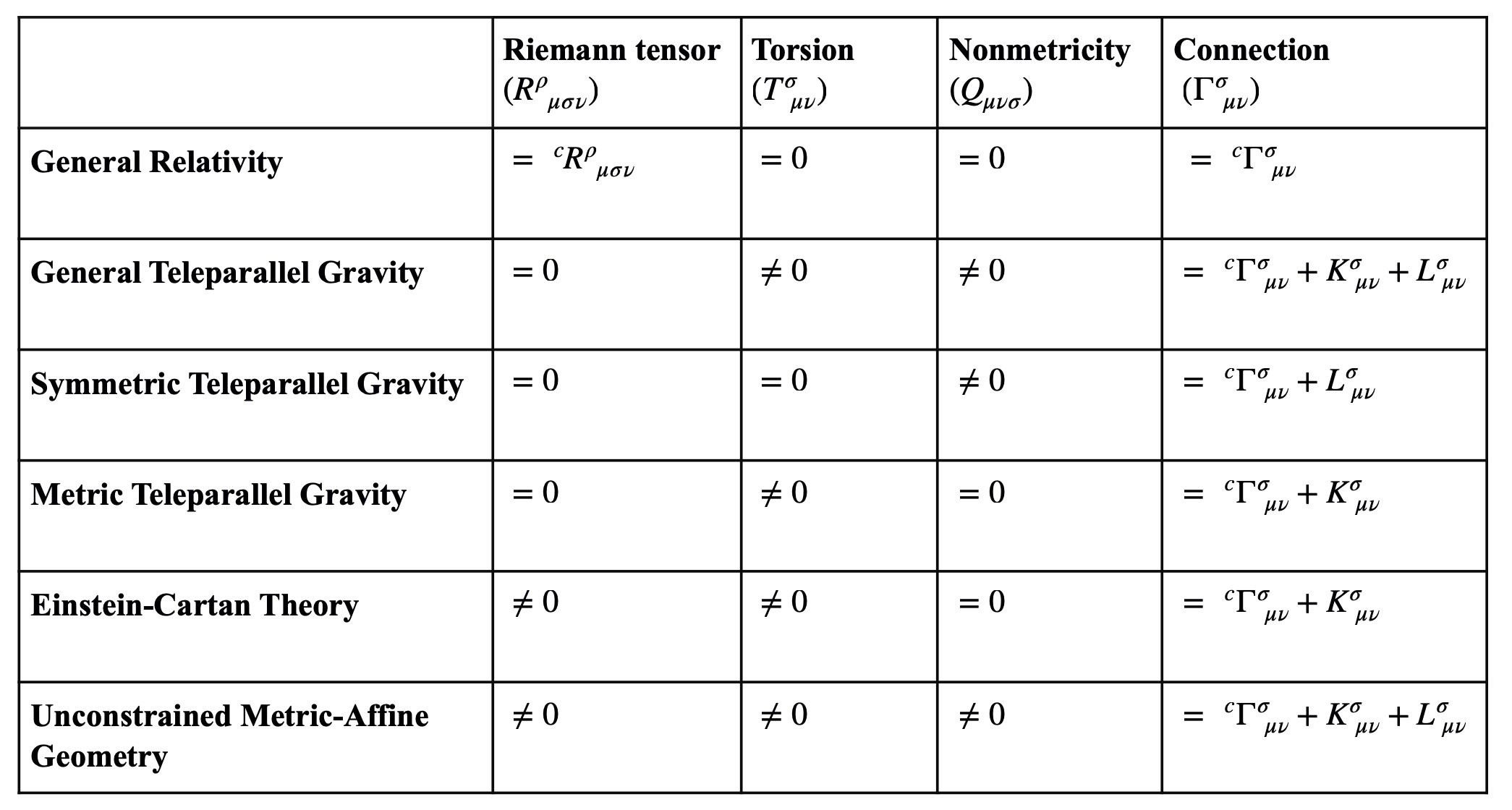} \caption{Schematic classification of gravitational theories according to the
geometric properties of spacetime. The table displays the presence
or absence of curvature (Riemann tensor $R^{\rho}{}_{\mu\sigma\nu}$),
torsion ($T^{\sigma}{}_{\mu\nu}$), and nonmetricity ($Q_{\mu\nu\sigma}$),
together with the corresponding form of the affine connection $\Gamma^{\sigma}{}_{\mu\nu}$.
Here $^{c}\Gamma^{\sigma}{}_{\mu\nu}$ denotes the Levi--Civita connection,
$K^{\sigma}{}_{\mu\nu}$ the contorsion tensor associated with torsion,
and $L^{\sigma}{}_{\mu\nu}$ the disformation tensor associated with
nonmetricity. General relativity is recovered when both torsion and
nonmetricity vanish. Einstein--Cartan theory allows for nonvanishing
torsion while preserving metric compatibility. Teleparallel theories
are characterized by the vanishing of the curvature, with gravitation
encoded entirely in torsion and/or nonmetricity. }
\label{fig:metric_affine_table} 
\end{figure}

\section{CONVENTIONS}

\label{sec:preliminaries}

In a metric-affine spacetime, the metric $g_{\mu\nu}$ and the affine
connection $\Gamma^{\sigma}{}_{\mu\nu}$ are treated as independent
geometric objects. The departure from Riemannian geometry is encoded
in two tensorial quantities: the nonmetricity 
\begin{equation}
Q_{\mu\nu\sigma}:=\nabla_{\mu}g_{\nu\sigma},\label{eq:def_nonmetricity}
\end{equation}
which measures the failure of the connection to preserve the metric
under parallel transport and the torsion 
\begin{equation}
T^{\sigma}{}_{\mu\nu}:=\Gamma^{\sigma}{}_{\mu\nu}-\Gamma^{\sigma}{}_{\nu\mu},\label{eq:def_torsion}
\end{equation}
which captures the antisymmetric part of the connection and governs the nonclosure of infinitesimal parallelograms.

Given the pair $(g_{\mu\nu},\Gamma^{\sigma}{}_{\mu\nu})$, a generic
affine connection may be uniquely decomposed into the Levi-Civita
connection plus tensorial corrections associated with torsion and
nonmetricity:\footnote{A detailed derivation of this decomposition is provided in Appendix~A.}
\begin{equation}
\Gamma^{\sigma}{}_{\mu\nu}={}^{c}\Gamma^{\sigma}{}_{\mu\nu}+K^{\sigma}{}_{\mu\nu}+L^{\sigma}{}_{\mu\nu}.\label{eq:Gamma_decomp}
\end{equation}
Here $^{c}\Gamma^{\sigma}{}_{\mu\nu}$ denotes the Levi-Civita (Christoffel)
connection \cite{RicciLeviCivita}, 
\begin{equation}
^{c}\Gamma^{\sigma}{}_{\mu\nu}=\frac{1}{2}\,g^{\sigma\lambda}\big(\partial_{\mu}g_{\nu\lambda}+\partial_{\nu}g_{\mu\lambda}-\partial_{\lambda}g_{\mu\nu}\big),\label{eq:LC_connection}
\end{equation}
$K^{\sigma}{}_{\mu\nu}$ is the contorsion tensor constructed from
the torsion, 
\begin{equation}
K^{\sigma}{}_{\mu\nu}=\frac{1}{2}\Big(T^{\sigma}{}_{\mu\nu}-T_{\mu}{}^{\sigma}{}_{\nu}-T_{\nu}{}^{\sigma}{}_{\mu}\Big),\label{eq:def_contorsion}
\end{equation}
and $L^{\sigma}{}_{\mu\nu}$ is the disformation tensor defined in
terms of the nonmetricity. With our index conventions, a convenient
and standard choice is\footnote{This definition ensures that $L^{\sigma}{}_{\mu\nu}$ is symmetric
in $\mu\nu$ whenever $Q_{\lambda\mu\nu}$ is symmetric in $\mu\nu$.
Adopting a different sign convention for $Q_{\mu\nu\sigma}$ would
correspondingly modify the signs in $L^{\sigma}{}_{\mu\nu}$.} 
\begin{equation}
L^{\sigma}{}_{\mu\nu}=\frac{1}{2}\Big(-\,Q^{\sigma}{}_{\mu\nu}+Q_{\mu}{}^{\sigma}{}_{\nu}+Q_{\nu}{}^{\sigma}{}_{\mu}\Big).\label{eq:def_disformation}
\end{equation}
Lowering the first index with the metric, these tensors obey useful 
symmetry properties 
\begin{align}
Q_{\sigma\mu\nu} & =Q_{\sigma\nu\mu}, & L_{\sigma\mu\nu} & =L_{\sigma\nu\mu},\label{eq:Q_L_sym}
\end{align}
together with the antisymmetry of the contorsion in its first two
indices, 
\begin{equation}
K_{\sigma\mu\nu}=-K_{\mu\sigma\nu}.\label{eq:K_antisym_12}
\end{equation}
Moreover, the contorsion tensor is related to the torsion through
the identities 
\begin{align}
K_{\mu\sigma\nu}-K_{\nu\sigma\mu} & =T_{\sigma\nu\mu},\label{eq:K_T_id1}\\
K_{\mu\nu\sigma}-K_{\mu\sigma\nu} & =T_{\mu\nu\sigma}.\label{eq:K_T_id2}
\end{align}
Throughout this work, the covariant derivative $\nabla_{\mu}$ is
defined using the full affine connection $\Gamma^{\sigma}{}_{\mu\nu}$.
In particular, 
\begin{equation}
\nabla_{\mu}g_{\nu\sigma}=\partial_{\mu}g_{\nu\sigma}-\Gamma^{\lambda}{}_{\nu\mu}g_{\lambda\sigma}-\Gamma^{\lambda}{}_{\sigma\mu}g_{\nu\lambda},\label{eq:covder_metric}
\end{equation}
which reproduces the definition of the nonmetricity tensor \eqref{eq:def_nonmetricity}.


\section{THE CURVATURE TENSORS AND THE BIANCHI IDENTITIES}

\label{sec:curvature_decomposition}

The Riemann tensor \cite{Riemann} of the generic connection is defined
by 
\begin{equation}
R^{\rho}{}_{\mu\sigma\nu}:=\partial_{\sigma}\Gamma^{\rho}{}_{\mu\nu}-\partial_{\nu}\Gamma^{\rho}{}_{\mu\sigma}+\Gamma^{\rho}{}_{\lambda\sigma}\Gamma^{\lambda}{}_{\mu\nu}-\Gamma^{\rho}{}_{\lambda\nu}\Gamma^{\lambda}{}_{\mu\sigma},\label{eq:def_Riemann_generic}
\end{equation}
and remains antisymmetric in its last two indices, 
\begin{equation}
R^{\rho}{}_{\mu\sigma\nu}=-R^{\rho}{}_{\mu\nu\sigma}.\label{eq:Riemann_last_antisym}
\end{equation}
In terms of the decomposition \eqref{eq:Gamma_decomp}, one can express
$R^{\rho}{}_{\mu\sigma\nu}$ as the Levi-Civita \cite{RicciLeviCivita}
curvature plus terms involving $K^{\sigma}{}_{\mu\nu}$ and $L^{\sigma}{}_{\mu\nu}$.
(A detailed derivation is provided in Appendix B.) We denote the Levi-Civita
curvature by $^{c}R^{\rho}{}_{\mu\sigma\nu}$. The Ricci tensor and scalar curvature are defined by the standard contractions 
\begin{equation}
R_{\mu\nu}:=R^{\rho}{}_{\mu\rho\nu},\qquad R:=g^{\mu\nu}R_{\mu\nu},\label{eq:Ricci_scalar_defs}
\end{equation}
and they can be written in terms of $^{c}R_{\mu\nu}$, $K^{\sigma}{}_{\mu\nu}$ and $L^{\sigma}{}_{\mu\nu}$ as recorded in Appendix B. A key technical input is the Ricci identity for a general connection acting on an
arbitrary tensor $T^{\sigma\rho\cdots}{}_{\lambda\kappa\cdots}$:\footnote{A proof is given in Appendix~C.}
\begin{align}
\big[\nabla_{\mu},\nabla_{\nu}\big]T^{\sigma\rho\cdots}{}_{\lambda\kappa\cdots} & =T^{\gamma}{}_{\mu\nu}\,\nabla_{\gamma}T^{\sigma\rho\cdots}{}_{\lambda\kappa\cdots}-\;R^{\sigma}{}_{\gamma\mu\nu}\,T^{\gamma\rho\cdots}{}_{\lambda\kappa\cdots}-\;R^{\rho}{}_{\gamma\mu\nu}\,T^{\sigma\gamma\cdots}{}_{\lambda\kappa\cdots}-\cdots\nonumber \\
 & \quad+\;R^{\gamma}{}_{\lambda\mu\nu}\,T^{\sigma\rho\cdots}{}_{\gamma\kappa\cdots}+\;R^{\gamma}{}_{\kappa\mu\nu}\,T^{\sigma\rho\cdots}{}_{\lambda\gamma\cdots}+\cdots.\label{eq:Ricci_identity_generic}
\end{align}
The first and second Bianchi identities take the forms \footnote{Detailed computations are given in Appendix~D.}
\begin{align}
R^{\lambda}{}_{\nu\rho\sigma}+R^{\lambda}{}_{\sigma\nu\rho}+R^{\lambda}{}_{\rho\sigma\nu} & -\nabla_{\nu}T^{\lambda}{}_{\sigma\rho}-\nabla_{\sigma}T^{\lambda}{}_{\rho\nu}-\nabla_{\rho}T^{\lambda}{}_{\nu\sigma}\nonumber \\
 & \quad+T^{\gamma}{}_{\sigma\rho}T^{\lambda}{}_{\gamma\nu}+T^{\gamma}{}_{\rho\nu}T^{\lambda}{}_{\gamma\sigma}+T^{\gamma}{}_{\nu\sigma}T^{\lambda}{}_{\gamma\rho}=0,\label{eq:first_Bianchi}\\[4pt]
\nabla_{\nu}R^{\mu}{}_{\lambda\rho\sigma}+\nabla_{\rho}R^{\mu}{}_{\lambda\sigma\nu}+\nabla_{\sigma}R^{\mu}{}_{\lambda\nu\rho} & +R^{\mu}{}_{\lambda\nu\gamma}T^{\gamma}{}_{\rho\sigma}+R^{\mu}{}_{\lambda\rho\gamma}T^{\gamma}{}_{\sigma\nu}+R^{\mu}{}_{\lambda\sigma\gamma}T^{\gamma}{}_{\nu\rho}=0.\label{eq:second_Bianchi}
\end{align}
A crucial qualitative difference from the Riemannian case is that, in general, one loses the symmetry under the exchange of index pairs,
\begin{equation}
R_{\mu\nu\rho\sigma}\neq R_{\rho\sigma\mu\nu},\label{eq:pair_exchange_failure}
\end{equation}
because metric compatibility and torsion-freeness are no longer assumed.


\section{A COMPACT FORM OF THE SECOND BIANCHI IDENTITY AND ITS TRACE}

\label{sec:identities}

For later convenience, we introduce the torsion-dependent and nonmetricity-dependent
combinations 
\begin{align}
X^{\mu}{}_{\lambda\nu\rho\sigma} & :=R^{\mu}{}_{\lambda\nu\gamma}T^{\gamma}{}_{\rho\sigma}+R^{\mu}{}_{\lambda\rho\gamma}T^{\gamma}{}_{\sigma\nu}+R^{\mu}{}_{\lambda\sigma\gamma}T^{\gamma}{}_{\nu\rho},\label{eq:def_X}\\
Y_{\gamma\lambda\nu\rho\sigma} & :=-\,R^{\mu}{}_{\lambda\rho\sigma}\,Q_{\nu\mu\gamma}-\,R^{\mu}{}_{\lambda\sigma\nu}\,Q_{\rho\mu\gamma}-\,R^{\mu}{}_{\lambda\nu\rho}\,Q_{\sigma\mu\gamma}.\label{eq:def_Y}
\end{align}
Then, the second Bianchi identity \ref{eq:second_Bianchi} can be written
in a compact form 
\begin{equation}
\nabla_{\nu}R_{\gamma\lambda\rho\sigma}+\nabla_{\rho}R_{\gamma\lambda\sigma\nu}+\nabla_{\sigma}R_{\gamma\lambda\nu\rho}+X_{\gamma\lambda\nu\rho\sigma}+Y_{\gamma\lambda\nu\rho\sigma}=0.\label{eq:second_Bianchiidentity}
\end{equation}
As expected, $X_{\gamma\lambda\nu\rho\sigma}=0$ when torsion vanishes,
and $Y_{\gamma\lambda\nu\rho\sigma}=0$ when the connection is metric
compatible. Tracing \eqref{eq:second_Bianchiidentity} with $g^{\nu\gamma}$
yields a generalized contracted Bianchi identity. After carefully
keeping track of the nonmetricity terms arising from raising indices,
one arrives at 
\begin{equation}
\nabla^{\gamma}R_{\gamma\lambda\rho\sigma}+\nabla_{\sigma}R_{\lambda\rho}-\nabla_{\rho}R_{\lambda\sigma}+\tilde{Y}_{\lambda\rho\sigma}+X_{\gamma\lambda}{}^{\gamma}{}_{\rho\sigma}=0,\label{eq:contracted_Bianchi_general}
\end{equation}
where we have defined 
\begin{equation}
\tilde{Y}_{\lambda\rho\sigma}:=Y_{\gamma\lambda}{}^{\gamma}{}_{\rho\sigma}-R_{\gamma\lambda\sigma\nu}\,Q_{\rho}{}^{\nu\gamma}-R_{\gamma\lambda\nu\rho}\,Q_{\sigma}{}^{\nu\gamma}.\label{eq:def_Ytilde}
\end{equation}
Similarly, it is convenient to rewrite the first Bianchi identity
\eqref{eq:first_Bianchi} as 
\begin{equation}
R^{\lambda}{}_{\nu\rho\sigma}+R^{\lambda}{}_{\sigma\nu\rho}+R^{\lambda}{}_{\rho\sigma\nu}+Z^{\lambda}{}_{\nu\rho\sigma}=0,\label{eq:first_Bianchi_compact}
\end{equation}
where 
\begin{align}
Z^{\lambda}{}_{\nu\rho\sigma} & :=-\nabla_{\nu}T^{\lambda}{}_{\sigma\rho}-\nabla_{\sigma}T^{\lambda}{}_{\rho\nu}-\nabla_{\rho}T^{\lambda}{}_{\nu\sigma}+T^{\gamma}{}_{\sigma\rho}T^{\lambda}{}_{\gamma\nu}+T^{\gamma}{}_{\rho\nu}T^{\lambda}{}_{\gamma\sigma}+T^{\gamma}{}_{\nu\sigma}T^{\lambda}{}_{\gamma\rho}.\label{eq:def_Z}
\end{align}


\section{WAVE-TYPE EQUATION FOR THE RIEMANN TENSOR}

\label{sec:dynamics}

We now derive a wave-type equation for the curvature in the presence
of torsion and nonmetricity. Taking $\nabla^{\nu}$ of \eqref{eq:second_Bianchiidentity}
and rearranging terms gives 
\begin{align}
\nabla^{\nu}\nabla_{\nu}R_{\gamma\lambda\rho\sigma}+\nabla^{\nu}\nabla_{\rho}R_{\gamma\lambda\sigma\nu}+\nabla^{\nu}\nabla_{\sigma}R_{\gamma\lambda\nu\rho}+\nabla^{\nu}\!\left(X_{\gamma\lambda\nu\rho\sigma}+Y_{\gamma\lambda\nu\rho\sigma}\right)=0.\label{eq:pre_wave_eq}
\end{align}
Introducing $\Box:=\nabla^{\nu}\nabla_{\nu}$ and commuting derivatives in the second and third terms yield 
\begin{align}
 & \Box R_{\gamma\lambda\rho\sigma}+\big[\nabla^{\nu},\nabla_{\rho}\big]R_{\gamma\lambda\sigma\nu}-\big[\nabla^{\nu},\nabla_{\sigma}\big]R_{\gamma\lambda\rho\nu}+\nabla_{\rho}\nabla^{\nu}R_{\gamma\lambda\sigma\nu}\nonumber \\
 & \quad\quad-\nabla_{\sigma}\nabla^{\nu}R_{\gamma\lambda\rho\nu}+\nabla^{\nu}\!\left(X_{\gamma\lambda\nu\rho\sigma}+Y_{\gamma\lambda\nu\rho\sigma}\right)=0.\label{eq:wave_eq_step1}
\end{align}
At this point, one may attempt to simplify the last-index divergences
$\nabla^{\nu}R_{\gamma\lambda\sigma\nu}$ and $\nabla^{\nu}R_{\gamma\lambda\rho\nu}$
using \eqref{eq:second_Bianchiidentity} and \eqref{eq:contracted_Bianchi_general}.
In a generic metric-affine geometry, however, these terms do not collapse
as neatly as in the Levi-Civita case; this is one of the technical
distinctions that makes the metric-affine curvature wave equation
structurally richer.

To proceed systematically, we evaluate the commutators in \eqref{eq:wave_eq_step1}
using the Ricci identity \eqref{eq:Ricci_identity_generic}. After
a straightforward but lengthy computation, one arrives at the following
wave-type equation (which is our main equation): 

\noindent\fbox{\begin{minipage}[t]{1\columnwidth - 2\fboxsep - 2\fboxrule}%
\begin{align}
\Box R_{\gamma\lambda\rho\sigma}+\nabla_{\rho}\nabla^{\nu}R_{\gamma\lambda\sigma\nu}-\nabla_{\sigma}\nabla^{\nu}R_{\gamma\lambda\rho\nu}+\nabla^{\nu}\!\left(X_{\gamma\lambda\nu\rho\sigma}+Y_{\gamma\lambda\nu\rho\sigma}\right)\nonumber \\
+T^{\nu k}{}_{\rho}\,\nabla_{\nu}R_{\gamma\lambda\sigma k}-T^{\nu k}{}_{\sigma}\,\nabla_{\nu}R_{\gamma\lambda\rho k}\nonumber \\
+R^{k}{}_{\gamma\rho}{}^{\nu}R_{k\lambda\sigma\nu}+R^{k}{}_{\lambda\rho}{}^{\nu}R_{\gamma k\sigma\nu}+R^{k}{}_{\sigma\rho}{}^{\nu}R_{\gamma\lambda k\nu}+R^{\kappa}\ _{\rho}R_{\gamma\lambda\sigma\kappa}\nonumber \\
-R^{k}{}_{\gamma\sigma}{}^{\nu}R_{k\lambda\rho\nu}-R^{k}{}_{\lambda\sigma}{}^{\nu}R_{\gamma k\rho\nu}-R^{k}{}_{\rho\sigma}{}^{\nu}R_{\gamma\lambda k\nu}-R^{\kappa}\ _{\sigma}R_{\gamma\lambda\rho\kappa} & =0,\label{eq:main_wave_equation_metric_affine}
\end{align}
\end{minipage}}\\
\\
where $R^{k}{}_{\ \rho}:=R^{k}{}_{\nu\rho}{}^{\nu}=R_{\rho}{}^{k}$
denotes the Ricci tensor with one index raised. Equation \eqref{eq:main_wave_equation_metric_affine}
is the metric-affine generalization of the familiar curvature wave
equation in Riemannian geometry: it reduces to the standard Levi--Civita
result when $T^{\sigma}{}_{\mu\nu}=0$ and $Q_{\mu\nu\sigma}=0$,
so that $X_{\gamma\lambda\nu\rho\sigma}=0=Y_{\gamma\lambda\nu\rho\sigma}$
and the additional torsion-driven transport terms vanish. Finally, for completeness, we explicitly record the dependence of $X$ and
$Y$ on torsion and nonmetricity: 
\begin{align}
X_{\gamma\lambda\nu\rho\sigma} & =g_{\mu\gamma}\,X^{\mu}{}_{\lambda\nu\rho\sigma}=g_{\mu\gamma}\Big(R^{\mu}{}_{\lambda\nu\kappa}T^{\kappa}{}_{\rho\sigma}+R^{\mu}{}_{\lambda\rho\kappa}T^{\kappa}{}_{\sigma\nu}+R^{\mu}{}_{\lambda\sigma\kappa}T^{\kappa}{}_{\nu\rho}\Big),\label{eq:X_explicit}\\
Y_{\gamma\lambda\nu\rho\sigma} & =-\,R^{\mu}{}_{\lambda\rho\sigma}Q_{\nu\mu\gamma}-\,R^{\mu}{}_{\lambda\sigma\nu}Q_{\rho\mu\gamma}-\,R^{\mu}{}_{\lambda\nu\rho}Q_{\sigma\mu\gamma}.\label{eq:Y_explicit}
\end{align}
In the next section, we analyze several instructive special cases
(Einstein spaces, teleparallel gravity \cite{AldrovandiPereira},
and Einstein-Cartan theory \cite{Cartan}), thereby illustrating how torsion and nonmetricity affect the propagation of curvature.\footnote{Teleparallel gravity and its subclasses are discussed in Appendix~E.}

\subsection{Wave equation in Einstein spaces}

\label{subsec:wave_Einstein_spaces}

In this subsection, we specialize the general curvature wave equation
\eqref{eq:main_wave_equation_metric_affine} to Einstein spaces, i.e., spacetimes whose Ricci tensor is proportional to the metric: 
\begin{equation}
R_{\mu\nu}=\lambda\,g_{\mu\nu},\label{eq:Einstein_space_def}
\end{equation}
where $\lambda$ is a constant.\footnote{In the purely Riemannian setting ($T^{\sigma}{}_{\mu\nu}=0=Q_{\mu\nu\sigma}$),
\eqref{eq:Einstein_space_def} implies $\lambda=\tfrac{2\Lambda}{n-2}$
in $n$ dimensions if the field equations are $R_{\mu\nu}-\tfrac{1}{2}Rg_{\mu\nu}+\Lambda g_{\mu\nu}=0$.
In the present metric-affine setting, we simply take \eqref{eq:Einstein_space_def}
as a geometric condition.} Using \eqref{eq:Einstein_space_def}, the Ricci terms in \eqref{eq:main_wave_equation_metric_affine}
simplify as 
\begin{equation}
R^{k}{}_{\ \rho}\,R_{\gamma\lambda\sigma k}-R^{k}{}_{\ \sigma}\,R_{\gamma\lambda\rho k}=\lambda\big(g^{k}{}_{\rho}R_{\gamma\lambda\sigma k}-g^{k}{}_{\sigma}R_{\gamma\lambda\rho k}\big)=\lambda\big(R_{\gamma\lambda\sigma\rho}-R_{\gamma\lambda\rho\sigma}\big).\label{eq:Einstein_Ricci_simplification_step}
\end{equation}
Since the Riemann tensor remains antisymmetric in its last two indices,
$R_{\gamma\lambda\sigma\rho}=-R_{\gamma\lambda\rho\sigma}$, the right-hand
side of \eqref{eq:Einstein_Ricci_simplification_step} becomes 
\begin{equation}
\lambda\big(R_{\gamma\lambda\sigma\rho}-R_{\gamma\lambda\rho\sigma}\big)=2\lambda\,R_{\gamma\lambda\sigma\rho}.\label{eq:Einstein_Ricci_simplification_final}
\end{equation}
Substituting \eqref{eq:Einstein_Ricci_simplification_final} into
the general wave equation \eqref{eq:main_wave_equation_metric_affine}
yields the Einstein-space form: 
\begin{align}
 & \Box R_{\gamma\lambda\rho\sigma}+\nabla_{\rho}\nabla^{\nu}R_{\gamma\lambda\sigma\nu}-\nabla_{\sigma}\nabla^{\nu}R_{\gamma\lambda\rho\nu}+\nabla^{\nu}\!\left(X_{\gamma\lambda\nu\rho\sigma}+Y_{\gamma\lambda\nu\rho\sigma}\right)\nonumber \\
 & \quad+T^{\nu k}{}_{\rho}\,\nabla_{\nu}R_{\gamma\lambda\sigma k}-T^{\nu k}{}_{\sigma}\,\nabla_{\nu}R_{\gamma\lambda\rho k}\nonumber \\
 & \quad+R^{k}{}_{\gamma\rho}{}^{\nu}R_{k\lambda\sigma\nu}+R^{k}{}_{\lambda\rho}{}^{\nu}R_{\gamma k\sigma\nu}+R^{k}{}_{\sigma\rho}{}^{\nu}R_{\gamma\lambda k\nu}\nonumber \\
 & \quad-R^{k}{}_{\gamma\sigma}{}^{\nu}R_{k\lambda\rho\nu}-R^{k}{}_{\lambda\sigma}{}^{\nu}R_{\gamma k\rho\nu}-R^{k}{}_{\rho\sigma}{}^{\nu}R_{\gamma\lambda k\nu}+2\lambda\,R_{\gamma\lambda\sigma\rho}=0.\label{eq:wave_equation_Einstein_space}
\end{align}
Equation \eqref{eq:wave_equation_Einstein_space} is the specialization
of the metric--affine curvature wave equation to Einstein spaces.
In the torsion-free and metric-compatible limit, the $X$- and $Y$-terms
vanish, and one recovers the familiar Riemannian curvature wave equation
in an Einstein background.

\subsection{Wave equation in the torsion- and nonmetricity-free case}

\label{subsec:wave_Riemannian_limit}

We now consider the Riemannian (Levi-Civita) limit in which both torsion
and nonmetricity vanish, 
\begin{equation}
T^{\mu}{}_{\nu\sigma}=0,\qquad Q_{\mu\nu\sigma}=0.\label{eq:TQ_zero}
\end{equation}
In this case, the connection reduces to the Levi-Civita one, and correspondingly,
the extra source terms introduced in \eqref{eq:def_X}--\eqref{eq:def_Y}
disappear: 
\begin{equation}
X_{\gamma\lambda\nu\rho\sigma}=0,\qquad Y_{\gamma\lambda\nu\rho\sigma}=0.\label{eq:XY_zero}
\end{equation}
The Riemann tensor then enjoys its standard Riemannian symmetries, namely, 
\begin{equation}
R_{\mu\nu\rho\sigma}=-R_{\nu\mu\rho\sigma}=-R_{\mu\nu\sigma\rho}=R_{\rho\sigma\mu\nu},\label{eq:Riemann_symmetries_Riemannian}
\end{equation}
together with the first Bianchi identity 
\begin{equation}
R^{k}{}_{\sigma\rho\nu}+R^{k}{}_{\rho\nu\sigma}+R^{k}{}_{\nu\sigma\rho}=0.\label{eq:first_Bianchi_Riemannian}
\end{equation}
Moreover, the contracted second Bianchi identity reduces to its familiar form, 
\begin{equation}
\nabla^{\gamma}R_{\gamma\lambda\rho\sigma}=\nabla_{\rho}R_{\lambda\sigma}-\nabla_{\sigma}R_{\lambda\rho}.\label{eq:contracted_Bianchi_Riemannian}
\end{equation}
Starting from the general wave-type equation \eqref{eq:main_wave_equation_metric_affine}
and imposing \eqref{eq:TQ_zero}--\eqref{eq:XY_zero}, we obtain
\begin{align}
 & \Box R_{\gamma\lambda\rho\sigma}+R^{k}{}_{\gamma\rho}{}^{\nu}R_{k\lambda\sigma\nu}+R^{k}{}_{\lambda\rho}{}^{\nu}R_{\gamma k\sigma\nu}+R^{k}{}_{\sigma\rho}{}^{\nu}R_{\gamma\lambda k\nu}+R^{k}{}_{\ \rho}\,R_{\gamma\lambda\sigma k}\nonumber \\
 & \quad-R^{k}{}_{\gamma\sigma}{}^{\nu}R_{k\lambda\rho\nu}-R^{k}{}_{\lambda\sigma}{}^{\nu}R_{\gamma k\rho\nu}-R^{k}{}_{\rho\sigma}{}^{\nu}R_{\gamma\lambda k\nu}-R^{k}{}_{\ \sigma}\,R_{\gamma\lambda\rho k}\nonumber \\
 & \quad+\nabla_{\rho}\nabla^{\nu}R_{\gamma\lambda\sigma\nu}-\nabla_{\sigma}\nabla^{\nu}R_{\gamma\lambda\rho\nu}=0.\label{eq:wave_Riemannian_step1}
\end{align}
Using the exchange symmetry $R_{\rho\sigma\mu\nu}=R_{\mu\nu\rho\sigma}$
in \eqref{eq:Riemann_symmetries_Riemannian} and the first Bianchi
identity \eqref{eq:first_Bianchi_Riemannian}, the purely quadratic
curvature terms in \eqref{eq:wave_Riemannian_step1} can be brought
to the standard symmetric form (see, e.g., the classic Lichnerowicz
identity). After these rearrangements, \eqref{eq:wave_Riemannian_step1}
may be written as 
\begin{align}
 & \Box R_{\gamma\lambda\rho\sigma}+2\,R_{k\lambda\sigma\nu}\,R^{k}{}_{\gamma\rho}{}^{\nu}+2\,R_{k\lambda\rho\nu}\,R^{k}{}_{\gamma\sigma}{}^{\nu}+R_{\gamma\lambda k\nu}\,R^{k\nu}{}_{\rho\sigma}\nonumber \\
 & \quad+R^{k}{}_{\ \rho}\,R_{\gamma\lambda\sigma k}-R^{k}{}_{\ \sigma}\,R_{\gamma\lambda\rho k}+\nabla_{\rho}\nabla^{\nu}R_{\gamma\lambda\sigma\nu}-\nabla_{\sigma}\nabla^{\nu}R_{\gamma\lambda\rho\nu}=0.\label{eq:wave_Riemannian_step2}
\end{align}
Finally, the remaining divergence terms are simplified using the contracted
Bianchi identity \eqref{eq:contracted_Bianchi_Riemannian}. In particular,
\begin{equation}
\nabla^{\nu}R_{\gamma\lambda\sigma\nu}=\nabla_{\lambda}R_{\gamma\sigma}-\nabla_{\gamma}R_{\lambda\sigma},\qquad\nabla^{\nu}R_{\gamma\lambda\rho\nu}=\nabla_{\lambda}R_{\gamma\rho}-\nabla_{\gamma}R_{\lambda\rho},\label{eq:div_Riemann_Riemannian}
\end{equation}
so that \eqref{eq:wave_Riemannian_step2} becomes 
\begin{align}
 & \Box R_{\gamma\lambda\rho\sigma}+2\,R_{k\lambda\sigma\nu}\,R^{k}{}_{\gamma\rho}{}^{\nu}+2\,R_{k\lambda\rho\nu}\,R^{k}{}_{\gamma\sigma}{}^{\nu}+R_{\gamma\lambda k\nu}\,R^{k\nu}{}_{\rho\sigma}\nonumber \\
 & \quad+R^{k}{}_{\ \rho}\,R_{\gamma\lambda\sigma k}-R^{k}{}_{\ \sigma}\,R_{\gamma\lambda\rho k}+\nabla_{\rho}\!\left(\nabla_{\lambda}R_{\gamma\sigma}-\nabla_{\gamma}R_{\lambda\sigma}\right)-\nabla_{\sigma}\!\left(\nabla_{\lambda}R_{\gamma\rho}-\nabla_{\gamma}R_{\lambda\rho}\right)=0.\label{eq:wave_Riemannian_final_compact}
\end{align}
Equivalently, expanding the last line yields the fully explicit form
\begin{align}
 & \Box R_{\gamma\lambda\rho\sigma}+2\,R_{k\lambda\sigma\nu}\,R^{k}{}_{\gamma\rho}{}^{\nu}+2\,R_{k\lambda\rho\nu}\,R^{k}{}_{\gamma\sigma}{}^{\nu}+R_{\gamma\lambda k\nu}\,R^{k\nu}{}_{\rho\sigma}+R^{k}{}_{\ \rho}\,R_{\gamma\lambda\sigma k}-R^{k}{}_{\ \sigma}\,R_{\gamma\lambda\rho k}\nonumber \\
 & \quad+\nabla_{\rho}\nabla_{\lambda}R_{\gamma\sigma}-\nabla_{\rho}\nabla_{\gamma}R_{\lambda\sigma}-\nabla_{\sigma}\nabla_{\lambda}R_{\gamma\rho}+\nabla_{\sigma}\nabla_{\gamma}R_{\lambda\rho}=0.\label{eq:wave_Riemannian_final_expanded}
\end{align}
Equation \eqref{eq:wave_Riemannian_final_expanded} is the standard curvature wave equation in Riemannian geometry that we discussed in the introduction section and serves as a consistency check of the
general metric-affine expression derived in the previous subsection.

\subsection{Wave equation in the nonmetricity-free case }

\label{subsec:wave_Q0_Tnonzero}

We next consider the metric-compatible case in which nonmetricity 
vanishes, 
\begin{equation}
Q_{\mu\nu\sigma}=0,\label{eq:Q_zero}
\end{equation}
while torsion is allowed to be nonzero, 
\begin{equation}
T^{\mu}{}_{\nu\sigma}\neq0.\label{eq:T_nonzero}
\end{equation}
Since $Q_{\mu\nu\sigma}=0$, the connection is metric compatible, and
the covariant derivative preserves the metric, $\nabla_{\mu}g_{\nu\sigma}=0$.
Consequently, indices may be raised and lowered freely with $g_{\mu\nu}$
without generating extra nonmetricity terms. In particular, the nonmetricity-dependent
tensor introduced in \eqref{eq:def_Y} vanishes identically: 
\begin{equation}
Y_{\gamma\lambda\nu\rho\sigma}=0.\label{eq:Y_zero_metric_compatible}
\end{equation}
Even though the torsion is present, the metric compatibility implies
that the curvature with all indices lowered is antisymmetric in its
first pair as well as in its second pair: 
\begin{equation}
R_{\mu\nu\rho\sigma}=-R_{\nu\mu\rho\sigma},\qquad R_{\mu\nu\rho\sigma}=-R_{\mu\nu\sigma\rho}.\label{eq:R_pair_antisym_Q0}
\end{equation}
However, in general, one still does not have the pair-exchange symmetry
\begin{equation}
R_{\mu\nu\rho\sigma}\neq R_{\rho\sigma\mu\nu},\label{eq:pair_exchange_fails_Q0}
\end{equation}
so the divergence of the Riemann tensor with respect to its last index
cannot, in general, be rewritten solely in terms of the Ricci tensor
as in the Levi--Civita case. This is one of the main structural differences
between torsionful (metric-compatible) geometries and Riemannian geometry.
Imposing \eqref{eq:Q_zero} (and hence \eqref{eq:Y_zero_metric_compatible})
in the general curvature wave equation \eqref{eq:main_wave_equation_metric_affine},
we obtain 
\begin{align}
 & \Box R_{\gamma\lambda\rho\sigma}+\nabla_{\rho}\nabla^{\nu}R_{\gamma\lambda\sigma\nu}-\nabla_{\sigma}\nabla^{\nu}R_{\gamma\lambda\rho\nu}+\nabla^{\nu}X_{\gamma\lambda\nu\rho\sigma}\nonumber \\
 & \quad+T^{\nu k}{}_{\rho}\,\nabla_{\nu}R_{\gamma\lambda\sigma k}-T^{\nu k}{}_{\sigma}\,\nabla_{\nu}R_{\gamma\lambda\rho k}\nonumber \\
 & \quad+R^{k}{}_{\gamma\rho}{}^{\nu}R_{k\lambda\sigma\nu}+R^{k}{}_{\lambda\rho}{}^{\nu}R_{\gamma k\sigma\nu}+R^{k}{}_{\sigma\rho}{}^{\nu}R_{\gamma\lambda k\nu}+R^{k}{}_{\ \rho}\,R_{\gamma\lambda\sigma k}\nonumber \\
 & \quad-R^{k}{}_{\gamma\sigma}{}^{\nu}R_{k\lambda\rho\nu}-R^{k}{}_{\lambda\sigma}{}^{\nu}R_{\gamma k\rho\nu}-R^{k}{}_{\rho\sigma}{}^{\nu}R_{\gamma\lambda k\nu}-R^{k}{}_{\ \sigma}\,R_{\gamma\lambda\rho k}=0.\label{eq:wave_equation_Q0}
\end{align}
Equation \eqref{eq:wave_equation_Q0} is the metric-compatible, torsionful
specialization of the general metric-affine curvature wave equation; it will be of particular interest when we further restrict to the Einstein-Cartan
theory in the next subsection. \footnote{For electromagnetic waves propagating on a background
with nonvanishing torsion, see \cite{Izaurieta}.}

\subsection{Wave equation in the torsion-free case }

\label{subsec:wave_T0_Qnonzero}

We finally consider the torsion-free sector of metric-affine geometry,
in which 
\begin{equation}
T^{\mu}{}_{\nu\sigma}=0,\label{eq:T_zero}
\end{equation}
while the nonmetricity is allowed to be nonvanishing, 
\begin{equation}
Q_{\mu\nu\sigma}\neq0.\label{eq:Q_nonzero}
\end{equation}
Since the torsion vanishes, the torsion-dependent tensor defined in
\eqref{eq:def_X} disappears identically, 
\begin{equation}
X_{\gamma\lambda\nu\rho\sigma}=0,\label{eq:X_zero_T0}
\end{equation}
whereas the nonmetricity-dependent tensor $Y_{\gamma\lambda\nu\rho\sigma}$
generally remains nonzero. In the absence of torsion, the curvature
tensor regains antisymmetry in its last two indices, 
\begin{equation}
R^{\mu}{}_{\nu\rho\sigma}=-R^{\mu}{}_{\nu\sigma\rho},\label{eq:R_last_pair_antisym_T0}
\end{equation}
but, due to nonmetricity, neither antisymmetry in the first index pair nor pair-exchange symmetry needs to hold. In particular, 
\begin{equation}
R_{\mu\nu\rho\sigma}\neq-R_{\nu\mu\rho\sigma},\qquad R_{\mu\nu\rho\sigma}\neq R_{\rho\sigma\mu\nu}.\label{eq:R_symmetry_failure_T0}
\end{equation}
As a consequence, raising and lowering indices with the metric produces
additional terms proportional to $Q_{\mu\nu\sigma}$, and the divergences of the Riemann tensor cannot be simplified using the standard Riemannian
identities.

Imposing \eqref{eq:T_zero} and \eqref{eq:X_zero_T0} in the general
curvature wave equation \eqref{eq:main_wave_equation_metric_affine},
we obtain 
\begin{align}
 & \Box R_{\gamma\lambda\rho\sigma}+\nabla_{\rho}\nabla^{\nu}R_{\gamma\lambda\sigma\nu}-\nabla_{\sigma}\nabla^{\nu}R_{\gamma\lambda\rho\nu}+\nabla^{\nu}Y_{\gamma\lambda\nu\rho\sigma}\nonumber \\
 & \quad+R^{k}{}_{\gamma\rho}{}^{\nu}R_{k\lambda\sigma\nu}+R^{k}{}_{\lambda\rho}{}^{\nu}R_{\gamma k\sigma\nu}+R^{k}{}_{\sigma\rho}{}^{\nu}R_{\gamma\lambda k\nu}+R^{k}{}_{\ \rho}\,R_{\gamma\lambda\sigma k}\nonumber \\
 & \quad-R^{k}{}_{\gamma\sigma}{}^{\nu}R_{k\lambda\rho\nu}-R^{k}{}_{\lambda\sigma}{}^{\nu}R_{\gamma k\rho\nu}-R^{k}{}_{\rho\sigma}{}^{\nu}R_{\gamma\lambda k\nu}-R^{k}{}_{\ \sigma}\,R_{\gamma\lambda\rho k}=0.\label{eq:wave_equation_T0}
\end{align}
Equation \eqref{eq:wave_equation_T0} is the torsion-free, nonmetricity
including the curvature wave equation. Compared to the metric-compatible
case \eqref{eq:wave_equation_Q0}, the only additional source term
is the divergence of $Y_{\gamma\lambda\nu\rho\sigma}$, which encodes the effect of nonmetricity on the propagation of curvature.
In the simultaneous limit $T^{\mu}{}_{\nu\sigma}=0$ and $Q_{\mu\nu\sigma}=0$,
this equation reduces smoothly to the standard Riemannian curvature
wave equation derived in Section~\ref{subsec:wave_Riemannian_limit}. In addition to the given constraints, namely, the
vanishing of torsion and the nonvanishing of nonmetricity, one may
further assume the vanishing of the Riemann tensor, which leads to
Symmetric Teleparallel Gravity (STGR).  Moreover,
although general relativity has no local degrees of freedom in three
dimensions, STGR does possess them \cite{Adak-Ozdemir}.

\section{CONCLUSIONS AND FURTHER DISCUSSIONS }

\label{sec:conclusions}

In Riemannian geometry \cite{Riemann} equipped with the Levi-Civita
connection \cite{RicciLeviCivita}, the curvature of spacetime is
not only a measure of gravitational interaction but also a dynamical
geometric object. As is well known, the Riemann tensor satisfies a
covariant wave equation whose source consists of terms quadratic in
curvature \cite{ChoquetBruhat2015}. This equation, which follows
from the Bianchi identities and the Ricci identity, provides a precise
geometric statement of the fact that curvature propagates: in this sense, Riemannian geometry is already wavy.

In this work, we have extended this familiar picture to the most general
metric-affine setting, where the affine connection possesses torsion 
and nonmetricity, in addition to curvature. Within this framework, we derived a wave-type equation satisfied by the Riemann tensor for
a generic connection. The resulting equation exhibits a rich structure: besides the usual curvature--curvature interaction terms known from
the Riemannian case, it contains additional contributions driven by
the torsion and nonmetricity. These appear both as algebraic couplings
to the curvature and as derivative (transport-like) terms, reflecting
the fact that torsion and nonmetricity modify the way geometric information
propagates across spacetime.

By analyzing several important limiting cases, we clarified how different
geometric structures affect the propagation of curvature. In the purely
Riemannian limit, where both the torsion and nonmetricity vanish,
our general expression reduces smoothly to the standard curvature
wave equation. In Einstein spaces, the equation simplifies further
and acquires an effective mass-like term proportional to the cosmological
constant. In metric-compatible but torsionful geometries \cite{AldrovandiPereira},
such as Einstein-Cartan theory \cite{Cartan}, torsion acts as an
additional source and transport mechanism for curvature waves. Conversely,
in torsion-free but nonmetric geometries \cite{BeltranJimenez}, the
influence of nonmetricity enters through explicit derivative couplings
that have no analog in Riemannian geometry.

Taken together, these results reinforce a unifying geometric viewpoint:
Gravity can be described equivalently in terms of curvature, 
torsion, or nonmetricity \cite{AldrovandiPereira}, but once these
structures are present, geometry itself becomes dynamic and wavy.
Curvature does not merely encode static geometric information; it propagates, interacts, and is influenced by the full affine structure
of spacetime.

Several directions for further investigation naturally emerge from
this work. First, it would be interesting to study the physical interpretation
of the additional torsion- and nonmetricity-induced terms in specific
models of metric-affine gravity, particularly in relation to spin,
dilation, and shear currents of matter.  One can find works in the literature along this direction \cite{Leh1,Leh2}
Second, linearizing the general
curvature wave equation around the appropriate backgrounds may shed
light on the propagation and polarization content of the geometric
waves beyond general relativity. Finally, the role of these curvature
wave equations in exact solutions, integrability properties, and conservation laws deserves further exploration.

We hope that the results presented here contribute to a deeper understanding
of the dynamical aspects of geometry in gravitational theories and
provide a useful bridge between the classical Riemannian picture and
its most general metric-affine extensions.

\section*{APPENDIX A: DECOMPOSITION OF A GENERIC AFFINE CONNECTION}

\label{app:connection_decomposition}

In this appendix, we derive the standard decomposition of a generic
affine connection into its Levi-Civita, torsional, and nonmetric
contributions. We assume that the spacetime metric $g_{\mu\nu}$ and
the affine connection $\Gamma^{\lambda}{}_{\mu\nu}$ are independent.

\subsection*{A.1 The nonmetricity and the torsion}

The nonmetricity tensor is defined by 
\begin{equation}
Q_{\sigma\mu\nu}:=\nabla_{\sigma}g_{\mu\nu}=\partial_{\sigma}g_{\mu\nu}-\Gamma^{\lambda}{}_{\mu\sigma}g_{\lambda\nu}-\Gamma^{\lambda}{}_{\nu\sigma}g_{\mu\lambda}.\label{eq:nonmetricity_def}
\end{equation}
Cyclic permutations of the indices yield 
\begin{align}
Q_{\mu\nu\sigma} & =\partial_{\mu}g_{\nu\sigma}-\Gamma^{\lambda}{}_{\nu\mu}g_{\lambda\sigma}-\Gamma^{\lambda}{}_{\sigma\mu}g_{\nu\lambda},\label{eq:Q_munu}\\
Q_{\nu\sigma\mu} & =\partial_{\nu}g_{\sigma\mu}-\Gamma^{\lambda}{}_{\sigma\nu}g_{\lambda\mu}-\Gamma^{\lambda}{}_{\mu\nu}g_{\sigma\lambda}.\label{eq:Q_nusig}
\end{align}
Combining \eqref{eq:nonmetricity_def}--\eqref{eq:Q_nusig}, we obtain
\begin{align}
Q_{\sigma\mu\nu}+Q_{\mu\nu\sigma}-Q_{\nu\sigma\mu}={} & \partial_{\sigma}g_{\mu\nu}+\partial_{\mu}g_{\nu\sigma}-\partial_{\nu}g_{\sigma\mu}\nonumber \\
 & -\Gamma^{\lambda}{}_{\mu\sigma}g_{\lambda\nu}-\Gamma^{\lambda}{}_{\nu\sigma}g_{\mu\lambda}-\Gamma^{\lambda}{}_{\nu\mu}g_{\lambda\sigma}\nonumber \\
 & -\Gamma^{\lambda}{}_{\sigma\mu}g_{\nu\lambda}+\Gamma^{\lambda}{}_{\sigma\nu}g_{\lambda\mu}+\Gamma^{\lambda}{}_{\mu\nu}g_{\sigma\lambda}.\label{eq:Q_combined}
\end{align}
Assuming a symmetric metric $g_{\mu\nu}=g_{\nu\mu}$, this simplifies
to 
\begin{align}
Q_{\sigma\mu\nu}+Q_{\mu\nu\sigma}-Q_{\nu\sigma\mu}={} & \partial_{\sigma}g_{\mu\nu}+\partial_{\mu}g_{\nu\sigma}-\partial_{\nu}g_{\sigma\mu}\nonumber \\
 & -\big(\Gamma^{\lambda}{}_{\mu\sigma}+\Gamma^{\lambda}{}_{\sigma\mu}\big)g_{\nu\lambda}+\big(\Gamma^{\lambda}{}_{\sigma\nu}-\Gamma^{\lambda}{}_{\nu\sigma}\big)g_{\mu\lambda}\nonumber \\
 & +\big(\Gamma^{\lambda}{}_{\mu\nu}-\Gamma^{\lambda}{}_{\nu\mu}\big)g_{\sigma\lambda}.\label{eq:Q_symmetric_metric}
\end{align}
The torsion tensor is defined as 
\begin{equation}
T^{\lambda}{}_{\mu\nu}:=2\Gamma^{\lambda}{}_{[\mu\nu]}=\Gamma^{\lambda}{}_{\mu\nu}-\Gamma^{\lambda}{}_{\nu\mu}.\label{eq:torsion_def}
\end{equation}
Using \eqref{eq:torsion_def}, equation \eqref{eq:Q_symmetric_metric}
becomes 
\begin{align}
Q_{\sigma\mu\nu}+Q_{\mu\nu\sigma}-Q_{\nu\sigma\mu}={} & \partial_{\sigma}g_{\mu\nu}+\partial_{\mu}g_{\nu\sigma}-\partial_{\nu}g_{\sigma\mu}\nonumber \\
 & -2\Gamma^{\lambda}{}_{\mu\sigma}g_{\nu\lambda}-T_{\nu\sigma\mu}+T_{\mu\sigma\nu}+T_{\sigma\mu\nu}.\label{eq:Q_with_torsion}
\end{align}

\subsection*{A.2 Extraction of the affine connection}

Contracting \eqref{eq:Q_with_torsion} with $\tfrac{1}{2}g^{\nu\beta}$
yields 
\begin{align}
\frac{1}{2}\big(Q_{\sigma\mu}{}^{\beta}+Q_{\mu}{}^{\beta}{}_{\sigma}-Q^{\beta}{}_{\sigma\mu}\big)={} & \frac{1}{2}g^{\nu\beta}\big(\partial_{\sigma}g_{\mu\nu}+\partial_{\mu}g_{\nu\sigma}-\partial_{\nu}g_{\sigma\mu}\big)\nonumber \\
 & -\Gamma^{\beta}{}_{\mu\sigma}+\frac{1}{2}\big(T_{\sigma\mu}{}^{\beta}+T_{\mu\sigma}{}^{\beta}-T^{\beta}{}_{\sigma\mu}\big).\label{eq:Gamma_extraction}
\end{align}
The first three terms on the right-hand side form the Levi-Civita
 connection, 
\begin{equation}
^{c}\Gamma^{\beta}{}_{\mu\sigma}:=\frac{1}{2}g^{\nu\beta}\big(\partial_{\sigma}g_{\mu\nu}+\partial_{\mu}g_{\nu\sigma}-\partial_{\nu}g_{\sigma\mu}\big),\label{eq:LC_def}
\end{equation}
which is symmetric in its lower indices. Using the symmetries 
\begin{equation}
Q_{\mu\nu\sigma}=Q_{\mu\sigma\nu},\qquad T_{\mu\nu\sigma}=-T_{\mu\sigma\nu},\label{eq:QT_symmetries}
\end{equation}
equation \eqref{eq:Gamma_extraction} can be solved for the full affine
connection: 
\begin{equation}
\Gamma^{\beta}{}_{\mu\sigma}={}^{c}\Gamma^{\beta}{}_{\mu\sigma}+\frac{1}{2}\big(Q^{\beta}{}_{\mu\sigma}-Q_{\sigma}{}^{\beta}{}_{\mu}-Q_{\mu}{}^{\beta}{}_{\sigma}\big)+\frac{1}{2}\big(T^{\beta}{}_{\mu\sigma}-T_{\sigma}{}^{\beta}{}_{\mu}-T_{\mu}{}^{\beta}{}_{\sigma}\big).\label{eq:Gamma_split_raw}
\end{equation}

\subsection*{A.3 The disformation and contorsion}

It is convenient to introduce the disformation tensor 
\begin{equation}
L^{\beta}{}_{\mu\sigma}:=\frac{1}{2}\big(Q^{\beta}{}_{\mu\sigma}-Q_{\sigma}{}^{\beta}{}_{\mu}-Q_{\mu}{}^{\beta}{}_{\sigma}\big),\label{eq:disformation_def}
\end{equation}
and the contorsion tensor 
\begin{equation}
K^{\beta}{}_{\mu\sigma}:=\frac{1}{2}\big(T^{\beta}{}_{\mu\sigma}-T_{\sigma}{}^{\beta}{}_{\mu}-T_{\mu}{}^{\beta}{}_{\sigma}\big).\label{eq:contorsion_def}
\end{equation}
With these definitions, the generic affine connection decomposes as
\begin{equation}
\boxed{\Gamma^{\beta}{}_{\mu\sigma}={}^{c}\Gamma^{\beta}{}_{\mu\sigma}+L^{\beta}{}_{\mu\sigma}+K^{\beta}{}_{\mu\sigma}.}\label{eq:Gamma_decomposition_final}
\end{equation}
Equation \eqref{eq:Gamma_decomposition_final} expresses the most
general affine connection as the sum of the Levi--Civita connection,
a purely nonmetric contribution, and a purely torsional contribution.

\subsection*{A.4 Properties of the torsion, non-metricity, contorsion, and the
disformation tensors}
\begin{itemize}
\item The Non-metricity tensor is symmetric in its last two indices: 
\begin{equation}
Q_{\sigma\mu\nu}=Q_{\sigma\nu\mu}.
\end{equation}
\item The disformation tensor is symmetric in its last two indices: 
\begin{equation}
L_{\sigma\mu\nu}=L_{\sigma\nu\mu}.
\end{equation}
\end{itemize}
PROOF: By definition, one has 
\begin{equation}
\begin{aligned}L_{\sigma\mu\nu} & =\frac{1}{2}\left(Q_{\sigma\mu\nu}-Q_{\mu\sigma\nu}-Q_{\nu\sigma\mu}\right)\\
 & =\frac{1}{2}\left(Q_{\sigma\nu\mu}-Q_{\nu\sigma\mu}-Q_{\mu\sigma\nu}\right)\\
 & =L_{\sigma\nu\mu}.
\end{aligned}
\end{equation}

\begin{itemize}
\item contorsion is antisymmetric in its first two indices: 
\begin{equation}
K_{\sigma\mu\nu}=-K_{\mu\sigma\nu}.
\end{equation}
\end{itemize}
PROOF: By definition, one has

\begin{equation}
\begin{aligned}K_{\sigma\mu\nu} & =\frac{1}{2}\left(T_{\sigma\mu\nu}-T_{\mu\sigma\nu}-T_{\nu\sigma\mu}\right)\\
 & =\frac{1}{2}\left(-T_{\sigma\nu\mu}+T_{\mu\nu\sigma}+T_{\nu\mu\sigma}\right),
\end{aligned}
\end{equation}
and equivalently, it becomes

\begin{equation}
\begin{aligned}K_{\sigma\mu\nu} & =\frac{1}{2}\left(-T_{\mu\sigma\nu}+T_{\sigma\mu\nu}+T_{\nu\mu\sigma}\right)\\
 & =-\frac{1}{2}\left(T_{\mu\sigma\nu}-T_{\sigma\mu\nu}-T_{\nu\mu\sigma}\right)\\
 & =-K_{\mu\sigma\nu}.
\end{aligned}
\end{equation}

\begin{itemize}
\item Difference of the contorsion tensors: 
\end{itemize}
\begin{equation}
K_{\mu\sigma\nu}-K_{\nu\sigma\mu}=-T_{\sigma\mu\nu}=T_{\sigma\nu\mu}.
\end{equation}

PROOF: The left-hand side reads 
\begin{equation}
\begin{aligned}K_{\mu\sigma\nu}-K_{\nu\sigma\mu} & =\frac{1}{2}\left(T_{\mu\sigma\nu}-T_{\sigma\mu\nu}-T_{\nu\mu\sigma}-T_{\nu\sigma\mu}+T_{\sigma\nu\mu}+T_{\mu\nu\sigma}\right)\\
 & =\frac{1}{2}\left(T_{\mu\sigma\nu}-T_{\sigma\mu\nu}-T_{\nu\mu\sigma}+T_{\nu\mu\sigma}-T_{\sigma\mu\nu}-T_{\mu\sigma\nu}\right)\\
 & =\frac{1}{2}\left(-T_{\sigma\mu\nu}-T_{\sigma\mu\nu}\right)\\
 & =-T_{\sigma\mu\nu}.
\end{aligned}
\end{equation}

\begin{itemize}
\item One more difference of the contorsion tensors can be written as: 
\end{itemize}
\begin{equation}
K_{\mu\nu\sigma}-K_{\mu\sigma\nu}=T_{\mu\nu\sigma}.
\end{equation}

PROOF: The left-hand side reads 
\begin{equation}
\begin{aligned}K_{\mu\nu\sigma}-K_{\mu\sigma\nu} & =\frac{1}{2}\left(-T_{\mu\sigma\nu}+T_{\sigma\mu\nu}+T_{\nu\mu\sigma}+T_{\mu\nu\sigma}-T_{\nu\mu\sigma}-T_{\sigma\mu\nu}\right)\\
 & =\frac{1}{2}\left(-T_{\mu\sigma\nu}+T_{\mu\nu\sigma}\right)\\
 & =\frac{1}{2}\left(T_{\mu\nu\sigma}+T_{\mu\nu\sigma}\right)\\
 & =T_{\mu\nu\sigma}.
\end{aligned}
\end{equation}

\begin{itemize}
\item Equality of the sums: 
\begin{equation}
K_{\mu\sigma\nu}+K_{\sigma\nu\mu}+K_{\nu\mu\sigma}=\frac{1}{2}\left(T_{\mu\sigma\nu}+T_{\sigma\nu\mu}+T_{\nu\mu\sigma}\right).
\end{equation}
\end{itemize}
PROOF: By definition, the left-hand side reads 
\begin{equation}
\begin{aligned}K_{\mu\sigma\nu}+K_{\sigma\nu\mu}+K_{\nu\mu\sigma}=\frac{1}{2} & \left(T_{\mu\sigma\nu}-T_{\sigma\mu\nu}-T_{\nu\mu\sigma}+T_{\sigma\nu\mu}-T_{\nu\sigma\mu}\right.\\
 & \left.-T_{\mu\sigma\nu}+T_{\nu\mu\sigma}-T_{\mu\nu\sigma}-T_{\sigma\nu\mu}\right)\\
= & \frac{1}{2}\left(-T_{\sigma\mu\nu}-T_{\nu\sigma\mu}-T_{\mu\nu\sigma}\right)\\
= & \frac{1}{2}\left(T_{\sigma\nu\mu}+T_{\nu\mu\sigma}+T_{\mu\sigma\nu}\right).
\end{aligned}
\end{equation}
Additionally, using the identity $K_{\mu\sigma\nu}-K_{\nu\sigma\mu}=T_{\sigma\nu\mu},$ we can express 
\begin{equation}
\begin{aligned}K_{\mu\sigma\nu}+K_{\sigma\nu\mu}+K_{\nu\mu\sigma} & =K_{\mu\sigma\nu}-K_{\nu\sigma\mu}+K_{\nu\mu\sigma}\\
 & =T_{\sigma\nu\mu}+K_{\nu\mu\sigma},
\end{aligned}
\end{equation}
which yields the following expression 
\begin{equation}
\begin{aligned}K_{\mu\sigma\nu}+K_{\sigma\nu\mu}+K_{\nu\mu\sigma} & =T_{\sigma\nu\mu}+\frac{1}{2}\left(T_{\nu\mu\sigma}-T_{\mu\nu\sigma}-T_{\sigma\nu\mu}\right)\\
 & =\frac{1}{2}\left(T_{\nu\mu\sigma}+T_{\sigma\nu\mu}+T_{\mu\sigma\nu}\right).
\end{aligned}
\end{equation}

\section*{APPENDIX B: THE CURVATURE TENSORS}

\label{app:curvature_tensors}

\subsection*{B.1 The Riemann tensor}

Given a generic affine connection $\Gamma^{\rho}{}_{\mu\nu}$, the
associated Riemann curvature tensor \cite{Riemann} is defined by
\begin{equation}
R^{\rho}{}_{\mu\sigma\nu}:=\partial_{\sigma}\Gamma^{\rho}{}_{\mu\nu}-\partial_{\nu}\Gamma^{\rho}{}_{\mu\sigma}+\Gamma^{\rho}{}_{\lambda\sigma}\Gamma^{\lambda}{}_{\mu\nu}-\Gamma^{\rho}{}_{\lambda\nu}\Gamma^{\lambda}{}_{\mu\sigma}.\label{eq:Riemann_def_appB}
\end{equation}
Throughout, we use the decomposition \eqref{eq:Gamma_decomposition_final}
derived in Appendix~A. Substituting \eqref{eq:Gamma_decomposition_final}
into \eqref{eq:Riemann_def_appB} and expanding, one may organize
the result as the Levi-Civita curvature, along with the terms involving
$K$ and $L$. The Levi-Civita Riemann tensor is 
\begin{equation}
^{c}R^{\rho}{}_{\mu\sigma\nu}:=\partial_{\sigma}\,{}^{c}\Gamma^{\rho}{}_{\mu\nu}-\partial_{\nu}\,{}^{c}\Gamma^{\rho}{}_{\mu\sigma}+{}^{c}\Gamma^{\rho}{}_{\lambda\sigma}\,{}^{c}\Gamma^{\lambda}{}_{\mu\nu}-{}^{c}\Gamma^{\rho}{}_{\lambda\nu}\,{}^{c}\Gamma^{\lambda}{}_{\mu\sigma}.\label{eq:LC_Riemann_appB}
\end{equation}
For compactness, it is convenient to introduce the distortion tensor
\begin{equation}
N^{\rho}{}_{\mu\nu}:=K^{\rho}{}_{\mu\nu}+L^{\rho}{}_{\mu\nu},\label{eq:distortion_def}
\end{equation}
which is a $(1,2)$ tensor. Then the curvature decomposition takes
the standard form 
\begin{equation}
R^{\rho}{}_{\mu\sigma\nu}={}^{c}R^{\rho}{}_{\mu\sigma\nu}+{}^{c}\nabla_{\sigma}N^{\rho}{}_{\mu\nu}-{}^{c}\nabla_{\nu}N^{\rho}{}_{\mu\sigma}+N^{\rho}{}_{\lambda\sigma}N^{\lambda}{}_{\mu\nu}-N^{\rho}{}_{\lambda\nu}N^{\lambda}{}_{\mu\sigma},\label{eq:Riemann_distortion_form}
\end{equation}
where $^{c}\nabla$ denotes the Levi--Civita covariant derivative.
Expanding $N^{\rho}{}_{\mu\nu}=K^{\rho}{}_{\mu\nu}+L^{\rho}{}_{\mu\nu}$
in \eqref{eq:Riemann_distortion_form}, we obtain an explicit expression
in terms of contorsion and disformation: 
\begin{align}
R^{\rho}{}_{\mu\sigma\nu}={} & ^{c}R^{\rho}{}_{\mu\sigma\nu}+{}^{c}\nabla_{\sigma}\!\left(K^{\rho}{}_{\mu\nu}+L^{\rho}{}_{\mu\nu}\right)-{}^{c}\nabla_{\nu}\!\left(K^{\rho}{}_{\mu\sigma}+L^{\rho}{}_{\mu\sigma}\right)\nonumber \\
 & \quad+\left(K^{\rho}{}_{\lambda\sigma}+L^{\rho}{}_{\lambda\sigma}\right)\left(K^{\lambda}{}_{\mu\nu}+L^{\lambda}{}_{\mu\nu}\right)-\left(K^{\rho}{}_{\lambda\nu}+L^{\rho}{}_{\lambda\nu}\right)\left(K^{\lambda}{}_{\mu\sigma}+L^{\lambda}{}_{\mu\sigma}\right).\label{eq:Riemann_KL_LC}
\end{align}
It is sometimes useful to rewrite the mixed torsion contribution using
the identity 
\begin{equation}
T^{\lambda}{}_{\nu\sigma}=\Gamma^{\lambda}{}_{\nu\sigma}-\Gamma^{\lambda}{}_{\sigma\nu}=K^{\lambda}{}_{\nu\sigma}-K^{\lambda}{}_{\sigma\nu},\label{eq:T_in_terms_of_K_appB}
\end{equation}
which follows from the definition of contorsion. With this, one may
equivalently present the decomposition in the form used in the main
text: 
\begin{align}
R^{\rho}{}_{\mu\sigma\nu}={} & ^{c}R^{\rho}{}_{\mu\sigma\nu}+\nabla_{\sigma}\!\left(K^{\rho}{}_{\mu\nu}+L^{\rho}{}_{\mu\nu}\right)-\nabla_{\nu}\!\left(K^{\rho}{}_{\mu\sigma}+L^{\rho}{}_{\mu\sigma}\right)\nonumber \\
 & \quad+\left(K^{\rho}{}_{\mu\lambda}+L^{\rho}{}_{\mu\lambda}\right)\left(K^{\lambda}{}_{\nu\sigma}-K^{\lambda}{}_{\sigma\nu}\right)\nonumber \\
 & \quad+\left(K^{\lambda}{}_{\mu\sigma}+L^{\lambda}{}_{\mu\sigma}\right)\left(K^{\rho}{}_{\lambda\nu}+L^{\rho}{}_{\lambda\nu}\right)-\left(K^{\lambda}{}_{\mu\nu}+L^{\lambda}{}_{\mu\nu}\right)\left(K^{\rho}{}_{\lambda\sigma}+L^{\rho}{}_{\lambda\sigma}\right),\label{eq:Riemann_maintext_form}
\end{align}
where $\nabla$ is the covariant derivative built from the full connection
$\Gamma^{\rho}{}_{\mu\nu}$. The two representations \eqref{eq:Riemann_KL_LC}
and \eqref{eq:Riemann_maintext_form} are equivalent; the former is
often cleaner for algebraic manipulations (it uses $^{c}\nabla$),
while the latter is convenient for keeping expressions manifestly
covariant with respect to the full connection.

Finally, irrespective of torsion and nonmetricity, the curvature defined
by \eqref{eq:Riemann_def_appB} remains antisymmetric in its last
two indices: 
\begin{equation}
R^{\rho}{}_{\mu\sigma\nu}=-R^{\rho}{}_{\mu\nu\sigma}.\label{eq:Riemann_last_antisym_appB}
\end{equation}

\subsubsection*{A quick check of curvature symmetries}

In the metric--affine case, we obtained the Riemann tensor in the
form 
\begin{align}
R^{\rho}{}_{\mu\nu\sigma}={} & ^{c}R^{\rho}{}_{\mu\nu\sigma}+\nabla_{\nu}\!\left(K^{\rho}{}_{\mu\sigma}+L^{\rho}{}_{\mu\sigma}\right)-\nabla_{\sigma}\!\left(K^{\rho}{}_{\mu\nu}+L^{\rho}{}_{\mu\nu}\right)\nonumber \\
 & \quad+\left(K^{\rho}{}_{\mu\lambda}+L^{\rho}{}_{\mu\lambda}\right)\left(K^{\lambda}{}_{\sigma\nu}-K^{\lambda}{}_{\nu\sigma}\right)\nonumber \\
 & \quad+\left(K^{\lambda}{}_{\mu\nu}+L^{\lambda}{}_{\mu\nu}\right)\left(K^{\rho}{}_{\lambda\sigma}+L^{\rho}{}_{\lambda\sigma}\right)-\left(K^{\lambda}{}_{\mu\sigma}+L^{\lambda}{}_{\mu\sigma}\right)\left(K^{\rho}{}_{\lambda\nu}+L^{\rho}{}_{\lambda\nu}\right).\label{eq:Riemann_metric_affine_form_symcheck}
\end{align}
By definition \eqref{eq:Riemann_def_appB}, the curvature is always
antisymmetric in its last two indices: 
\begin{equation}
R^{\rho}{}_{\mu\nu\sigma}=-R^{\rho}{}_{\mu\sigma\nu}.\label{eq:R_last_pair_antisym_symcheck}
\end{equation}
The symmetry under the exchange of the first index pair is subtler.
In the most general metric--affine case ($Q_{\alpha\mu\nu}\neq0$), lowering indices involves the full connection and does not commute
with covariant differentiation; consequently, $R_{\mu\nu\rho\sigma}$
need not be antisymmetric in $\mu,\nu$. More explicitly, one generally
has 
\begin{equation}
R_{\mu\nu\rho\sigma}\neq-R_{\nu\mu\rho\sigma},\label{eq:first_pair_no_antisym_generic}
\end{equation}
and the pair-exchange symmetry also fails in general, 
\begin{equation}
R_{\mu\nu\rho\sigma}\neq R_{\rho\sigma\mu\nu}.\label{eq:pair_exchange_no_generic}
\end{equation}
On the other hand, in the metric-compatible case ($Q_{\alpha\mu\nu}=0$),
one can lower the indices without generating extra nonmetricity terms.
Then the curvature with all indices lowered regains antisymmetry in
the first pair, 
\begin{equation}
Q_{\alpha\mu\nu}=0\qquad\Longrightarrow\qquad R_{\mu\nu\rho\sigma}=-R_{\nu\mu\rho\sigma},\label{eq:first_pair_antisym_metric_compatible}
\end{equation}
while the pair-exchange symmetry $R_{\mu\nu\rho\sigma}=R_{\rho\sigma\mu\nu}$
may still fail when torsion is present.

\subsection*{B2. The Ricci tensor}

The Ricci tensor is defined by the contraction of the first and third
indices, 
\begin{equation}
R_{\mu\nu}:=R^{\sigma}{}_{\mu\sigma\nu},\qquad{}^{c}R_{\mu\nu}:={}^{c}R^{\sigma}{}_{\mu\sigma\nu}.\label{eq:Ricci_def_appB}
\end{equation}
Contracting \eqref{eq:Riemann_maintext_form} accordingly, one obtains
\begin{align}
R_{\mu\nu}={} & ^{c}R_{\mu\nu}+\nabla_{\sigma}\!\left(K^{\sigma}{}_{\mu\nu}+L^{\sigma}{}_{\mu\nu}\right)-\nabla_{\nu}\!\left(K^{\sigma}{}_{\mu\sigma}+L^{\sigma}{}_{\mu\sigma}\right)\nonumber \\
 & \quad+(K^{\lambda}{}_{\mu\sigma}+L^{\lambda}{}_{\mu\sigma})\left(K^{\sigma}{}_{\nu\lambda}+L^{\sigma}{}_{\nu\lambda}\right)-(K^{\lambda}{}_{\mu\nu}+L^{\lambda}{}_{\mu\nu})\left(K^{\sigma}{}_{\lambda\sigma}+L^{\sigma}{}_{\lambda\sigma}\right).\label{eq:Ricci_KL_final}
\end{align}
Here we used that the disformation tensor is symmetric in its last
two indices, $L^{\rho}{}_{\mu\nu}=L^{\rho}{}_{\nu\mu}$, while the
contorsion satisfies $K_{\rho\mu\nu}=-K_{\mu\rho\nu}$.

\subsection*{B3. The scalar curvature}

The scalar curvature is $R=g^{\mu\nu}R_{\mu\nu}$. Writing also $^{c}R=g^{\mu\nu}{}^{c}R_{\mu\nu}$,
we find from \eqref{eq:Ricci_KL_final} that: 
\begin{align}
R={} & ^{c}R+g^{\mu\nu}\nabla_{\sigma}\!\left(K^{\sigma}{}_{\mu\nu}+L^{\sigma}{}_{\mu\nu}\right)-\nabla_{\mu}\!\left(K^{\sigma}{}_{\mu\sigma}+L^{\sigma}{}_{\mu\sigma}\right)\nonumber \\
 & \quad+K^{\lambda\mu\sigma}K_{\sigma\mu\lambda}+L^{\lambda\mu\sigma}L_{\sigma\mu\lambda}+K^{\mu\lambda}{}_{\mu}K^{\sigma}{}_{\lambda\sigma}+K^{\mu\lambda}{}_{\mu}L^{\sigma}{}_{\lambda\sigma}-L^{\lambda\mu}{}_{\mu}K^{\sigma}{}_{\lambda\sigma}-L^{\lambda\mu}{}_{\mu}L^{\sigma}{}_{\lambda\sigma}.\label{eq:Scalar_curvature_KL_final}
\end{align}
Equation \eqref{eq:Scalar_curvature_KL_final} is the scalar curvature
in terms of Levi--Civita curvature, plus the torsion and the nonmetricity
contributions. Note that in the generic case one has $\nabla_{\sigma}g^{\mu\nu}\neq0$;
hence, the divergence terms in \eqref{eq:Scalar_curvature_KL_final}
should be kept as written.

\subsubsection*{Metric-compatible limit}

When the connection is metric compatible ($Q_{\alpha\mu\nu}=0$),
the disformation vanishes, $L^{\rho}{}_{\mu\nu}=0$, and \eqref{eq:Scalar_curvature_KL_final}
reduces to 
\begin{align}
R={} & ^{c}R+g^{\mu\nu}\nabla_{\sigma}K^{\sigma}{}_{\mu\nu}-\nabla_{\mu}K^{\sigma}{}_{\mu\sigma}+K^{\lambda\mu\sigma}K_{\sigma\mu\lambda}+K^{\mu\lambda}{}_{\mu}K^{\sigma}{}_{\lambda\sigma}.\label{eq:Scalar_curvature_metric_compatible_step1}
\end{align}
Using $Q_{\alpha\mu\nu}=0$ one may move the metric through the covariant
derivative, so $g^{\mu\nu}\nabla_{\sigma}K^{\sigma}{}_{\mu\nu}=\nabla_{\sigma}K^{\sigma\mu}{}_{\mu}$,
and therefore 
\begin{equation}
R=^{c}R+2\,\nabla_{\sigma}\!K^{\sigma\mu}{}_{\mu}+K^{\lambda\mu\sigma}K_{\sigma\mu\lambda}+K^{\mu\lambda}{}_{\mu}K^{\sigma}{}_{\lambda\sigma}.\label{eq:Scalar_curvature_metric_compatible_final}
\end{equation}

\section*{APPENDIX C: THE RICCI IDENTITY FOR THE COVARIANT DERIVATIVES}

\label{app:ricci_identity}

In this appendix, we derive the commutator (Ricci) identity for a
generic affine connection with torsion and possibly nonmetricity.
The nonmetricity affects how the indices are raised and lowered, but
the commutator formula itself is determined by the torsion and curvature.

\subsection*{C.1 Warm-up: action on a scalar}

Let $f$ be a scalar field. Since $\nabla_{\mu}f=\partial_{\mu}f$,
we compute 
\begin{align}
[\nabla_{\mu},\nabla_{\nu}]f & =\nabla_{\mu}\nabla_{\nu}f-\nabla_{\nu}\nabla_{\mu}f\nonumber \\
 & =\nabla_{\mu}(\partial_{\nu}f)-\nabla_{\nu}(\partial_{\mu}f)\nonumber \\
 & =\partial_{\mu}\partial_{\nu}f-\Gamma^{\lambda}{}_{\nu\mu}\partial_{\lambda}f-\partial_{\nu}\partial_{\mu}f+\Gamma^{\lambda}{}_{\mu\nu}\partial_{\lambda}f\nonumber \\
 & =\left(\Gamma^{\lambda}{}_{\mu\nu}-\Gamma^{\lambda}{}_{\nu\mu}\right)\partial_{\lambda}f=T^{\lambda}{}_{\mu\nu}\,\nabla_{\lambda}f,\label{eq:comm_scalar}
\end{align}
where the torsion tensor is $T^{\lambda}{}_{\mu\nu}:=\Gamma^{\lambda}{}_{\mu\nu}-\Gamma^{\lambda}{}_{\nu\mu}$.
Thus, unlike the Levi-Civita case, the commutator does not vanish
even for scalars.

\subsection*{C.2 Action on a covector and a vector}

For a covector $V_{\sigma}$ one finds 
\begin{equation}
[\nabla_{\mu},\nabla_{\nu}]V_{\sigma}=T^{\lambda}{}_{\mu\nu}\,\nabla_{\lambda}V_{\sigma}+R^{\lambda}{}_{\sigma\nu\mu}\,V_{\lambda},\label{eq:comm_covector}
\end{equation}
where the Riemann tensor $R^{\lambda}{}_{\sigma\nu\mu}$ is defined
by \eqref{eq:Riemann_def_appB}. For a vector $V^{\sigma}$ one similarly
obtains 
\begin{equation}
[\nabla_{\mu},\nabla_{\nu}]V^{\sigma}=T^{\lambda}{}_{\mu\nu}\,\nabla_{\lambda}V^{\sigma}-R^{\sigma}{}_{\lambda\nu\mu}\,V^{\lambda}.\label{eq:comm_vector}
\end{equation}
The sign difference between \eqref{eq:comm_covector} and \eqref{eq:comm_vector}
is the usual one: curvature acts with the opposite sign on the covariant
versus contravariant indices.

\subsection*{C.3 General Ricci identity}

For a general tensor field $A^{\sigma\rho\cdots}{}_{\lambda k\cdots}$, repeated application of the Leibniz rule together with the results
of Subsections~C.1 and~C.2 yields the Ricci (commutator) identity
for a generic affine connection with torsion: 
\begin{align}
[\nabla_{\mu},\nabla_{\nu}]\,A^{\sigma\rho\cdots}{}_{\lambda k\cdots}={} & T^{\gamma}{}_{\mu\nu}\,\nabla_{\gamma}A^{\sigma\rho\cdots}{}_{\lambda k\cdots}\nonumber \\
 & -R^{\sigma}{}_{\gamma\nu\mu}\,A^{\gamma\rho\cdots}{}_{\lambda k\cdots}-R^{\rho}{}_{\gamma\nu\mu}\,A^{\sigma\gamma\cdots}{}_{\lambda k\cdots}-\cdots\nonumber \\
 & +R^{\gamma}{}_{\lambda\nu\mu}\,A^{\sigma\rho\cdots}{}_{\gamma k\cdots}+R^{\gamma}{}_{k\nu\mu}\,A^{\sigma\rho\cdots}{}_{\lambda\gamma\cdots}+\cdots.\label{eq:Ricci_identity_general}
\end{align}
Here, each contravariant index contributes a curvature term with a
minus sign, while each covariant index contributes a curvature term
with a plus sign; the ellipsis indicates the obvious continuation
to all remaining indices of the tensor.

Equation~\eqref{eq:Ricci_identity_general} is the form used throughout
the main text. In the torsion-free limit, $T^{\gamma}{}_{\mu\nu}=0$,
it reduces to the standard Ricci identity of Riemannian geometry.

\section*{APPENDIX D: BIANCHI IDENTITIES}

In this appendix, we derive the generalized Bianchi identities for
a generic affine connection with torsion. The derivation is based
on the Jacobi identity satisfied by covariant derivatives.

\subsection*{D.1 Jacobi identity}

For any vector field $V^{\mu}$, the covariant derivatives satisfy
\begin{equation}
\left[\left[\nabla_{\sigma},\nabla_{\rho}\right],\nabla_{\nu}\right]V^{\mu}+\left[\left[\nabla_{\rho},\nabla_{\nu}\right],\nabla_{\sigma}\right]V^{\mu}+\left[\left[\nabla_{\nu},\nabla_{\sigma}\right],\nabla_{\rho}\right]V^{\mu}=0.\label{eq:Jacobi_identity}
\end{equation}
This identity will lead simultaneously to the first and second Bianchi identities.

\subsection*{D.2 Evaluation of the nested commutators}

We begin by evaluating the first term in \eqref{eq:Jacobi_identity}:
\begin{equation}
\left[\left[\nabla_{\sigma},\nabla_{\rho}\right],\nabla_{\nu}\right]V^{\mu}=\left[\nabla_{\sigma},\nabla_{\rho}\right]\nabla_{\nu}V^{\mu}-\nabla_{\nu}\left[\nabla_{\sigma},\nabla_{\rho}\right]V^{\mu}.
\end{equation}
Using the Ricci identity derived in Appendix~C, one finds 
\begin{equation}
\begin{aligned}\left[\left[\nabla_{\sigma},\nabla_{\rho}\right],\nabla_{\nu}\right]V^{\mu}={} & T^{\lambda}{}_{\sigma\rho}\nabla_{\lambda}\nabla_{\nu}V^{\mu}+R^{\lambda}{}_{\nu\rho\sigma}\nabla_{\lambda}V^{\mu}-R^{\mu}{}_{\lambda\rho\sigma}\nabla_{\nu}V^{\lambda}\\
 & -\nabla_{\nu}\!\left(T^{\lambda}{}_{\sigma\rho}\nabla_{\lambda}V^{\mu}-R^{\mu}{}_{\lambda\rho\sigma}V^{\lambda}\right).
\end{aligned}
\end{equation}
Expanding the last term and regrouping yields 
\begin{equation}
\begin{aligned}\left[\left[\nabla_{\sigma},\nabla_{\rho}\right],\nabla_{\nu}\right]V^{\mu}={} & T^{\lambda}{}_{\sigma\rho}\left[\nabla_{\lambda},\nabla_{\nu}\right]V^{\mu}+\left(R^{\lambda}{}_{\nu\rho\sigma}-\nabla_{\nu}T^{\lambda}{}_{\sigma\rho}\right)\nabla_{\lambda}V^{\mu}\\
 & +V^{\lambda}\nabla_{\nu}R^{\mu}{}_{\lambda\rho\sigma}.
\end{aligned}
\end{equation}
Applying the Ricci identity once more, we arrive at 
\begin{equation}
\begin{aligned}\left[\left[\nabla_{\sigma},\nabla_{\rho}\right],\nabla_{\nu}\right]V^{\mu}={} & \left(R^{\lambda}{}_{\nu\rho\sigma}-\nabla_{\nu}T^{\lambda}{}_{\sigma\rho}+T^{\gamma}{}_{\sigma\rho}T^{\lambda}{}_{\gamma\nu}\right)\nabla_{\lambda}V^{\mu}\\
 & +\left(\nabla_{\nu}R^{\mu}{}_{\lambda\rho\sigma}-R^{\mu}{}_{\lambda\nu\gamma}T^{\gamma}{}_{\sigma\rho}\right)V^{\lambda}.
\end{aligned}
\label{eq:first_nested}
\end{equation}
The remaining two nested commutators are obtained by cyclic permutations
of $(\sigma,\rho,\nu)$.

\subsection*{D.3 Generalized Bianchi identities}

Substituting the three cyclic permutations of \eqref{eq:first_nested}
into the Jacobi identity \eqref{eq:Jacobi_identity} and collecting
terms proportional to $\nabla_{\lambda}V^{\mu}$ and $V^{\lambda}$
separately, one obtains 
\begin{align}
 & R^{\lambda}{}_{\nu\rho\sigma}+R^{\lambda}{}_{\sigma\nu\rho}+R^{\lambda}{}_{\rho\sigma\nu}-\nabla_{\nu}T^{\lambda}{}_{\sigma\rho}-\nabla_{\sigma}T^{\lambda}{}_{\rho\nu}-\nabla_{\rho}T^{\lambda}{}_{\nu\sigma}\nonumber \\
 & \qquad+T^{\gamma}{}_{\sigma\rho}T^{\lambda}{}_{\gamma\nu}+T^{\gamma}{}_{\rho\nu}T^{\lambda}{}_{\gamma\sigma}+T^{\gamma}{}_{\nu\sigma}T^{\lambda}{}_{\gamma\rho}=0,\label{eq:first_Bianchi_full}
\end{align}
and 
\begin{align}
 & \nabla_{\nu}R^{\mu}{}_{\lambda\rho\sigma}+\nabla_{\rho}R^{\mu}{}_{\lambda\sigma\nu}+\nabla_{\sigma}R^{\mu}{}_{\lambda\nu\rho}\nonumber \\
 & \qquad-R^{\mu}{}_{\lambda\nu\gamma}T^{\gamma}{}_{\sigma\rho}-R^{\mu}{}_{\lambda\sigma\gamma}T^{\gamma}{}_{\rho\nu}-R^{\mu}{}_{\lambda\rho\gamma}T^{\gamma}{}_{\nu\sigma}=0.\label{eq:second_Bianchi_full}
\end{align}

\subsection*{D.4 Compact antisymmetric form}

Using antisymmetrization brackets, the first Bianchi identity \eqref{eq:first_Bianchi_full}
can be written compactly as 
\begin{equation}
R^{\lambda}{}_{[\nu\rho\sigma]}=-\nabla_{[\nu}T^{\lambda}{}_{\rho\sigma]}+T^{\gamma}{}_{[\nu\rho}T^{\lambda}{}_{\sigma]\gamma}.\label{eq:first_Bianchi_compact}
\end{equation}
Similarly, the second Bianchi identity \eqref{eq:second_Bianchi_full}
takes the compact form 
\begin{equation}
\nabla_{[\nu}R^{\mu}{}_{|\lambda|\rho\sigma]}=R^{\mu}{}_{\lambda\gamma[\nu}T^{\gamma}{}_{\rho\sigma]}.\label{eq:second_Bianchi_compact}
\end{equation}
In the torsion-free limit $T^{\lambda}{}_{\mu\nu}=0$, these identities
reduce to the familiar Bianchi identities of Riemannian geometry.

\section*{APPENDIX E: TELEPARALLEL GRAVITY}

Recall the generic expression for the Riemann tensor 
\begin{equation}
\begin{aligned}R_{\ \mu\sigma\nu}^{\rho}=\  & ^{c}R_{\ \mu\sigma\nu}^{\rho}+\nabla_{\sigma}K_{\ \mu\nu}^{\rho}+\nabla_{\sigma}L_{\ \mu\nu}^{\rho}-\nabla_{\nu}K_{\ \mu\sigma}^{\rho}-\nabla_{\nu}L^{\rho}{}_{\mu\sigma}\\
 & +\left(K_{\ \mu\lambda}^{\rho}+L_{\ \mu\lambda}^{\rho}\right)\left(K^{\lambda}{}_{\nu\sigma}-K^{\lambda}{}_{\sigma\nu}\right)\\
 & +\left(K^{\lambda}{}_{\mu\sigma}+L^{\lambda}{}_{\mu\sigma}\right)\left(K^{\rho}{}_{\lambda\nu}+L^{\rho}{}_{\lambda\nu}\right)\\
 & -\left(K^{\lambda}{}_{\mu\nu}+L^{\lambda}{}_{\mu\nu}\right)\left(K^{\rho}{}_{\lambda\sigma}+L^{\rho}{}_{\lambda\sigma}\right).
\end{aligned}
\end{equation}
Using the splitting of the generic connection \eqref{eq:Gamma_decomposition_final}
and introducing the distortion tensor $D^{\mu}{}_{\nu\sigma}$ 
\begin{gather}
D^{\mu}{}_{\nu\sigma}:=\Gamma^{\mu}{}_{\nu\sigma}-^{c}\Gamma^{\mu}{}_{\nu\sigma}=K^{\mu}{}_{\nu\sigma}+L^{\mu}{}_{\nu\sigma},
\end{gather}
we can rewrite the Riemann tensor as follows 
\begin{equation}
\begin{aligned}R_{\ \mu\sigma\nu}^{\rho}=\  & ^{c}R_{\ \mu\sigma\nu}^{\rho}+\nabla_{\sigma}D_{\ \mu\nu}^{\rho}-\nabla_{\nu}D_{\ \mu\sigma}^{\rho}+D_{\ \mu\lambda}^{\rho}\left(K^{\lambda}{}_{\nu\sigma}-K^{\lambda}{}_{\sigma\nu}\right)\\
 & +D^{\lambda}{}_{\mu\sigma}D^{\rho}{}_{\lambda\nu}-D^{\lambda}{}_{\mu\nu}D^{\rho}{}_{\lambda\sigma}.
\end{aligned}
\end{equation}
Or equivalently, one arrives at 
\begin{equation}
\begin{array}{rc}
 & R_{\ \mu\sigma\nu}^{\rho}=\nabla_{\sigma}D_{\ \mu\nu}^{\rho}-\nabla_{\nu}D_{\ \mu\sigma}^{\rho}+D_{\ \mu\sigma}^{\lambda}D_{\ \lambda\nu}^{\rho}\\
 & \mathbf{\mathbf{\ \ \ \ \ }\mathbf{\ \ \ \ \ }\mathbf{\ \ \ \ \ }\ \ \ \ \ }\mathbf{\ \ \ \ \ }\mathbf{\ \ \ \ \ }-D^{\lambda}{}_{\mu\nu}D^{\rho}{}_{\lambda\sigma}+D_{\ \mu\lambda}^{\rho}\left(K^{\lambda}{}_{\nu\sigma}+L^{\lambda}{}_{\nu\sigma}-L^{\lambda}{}_{\nu\sigma}-K^{\lambda}{}_{\sigma\nu}\right),
\end{array}
\end{equation}
where one can use the symmetry $L^{\lambda}{}_{v\sigma}=L_{\ \sigma v}^{\lambda}$
in the last equation. Then the last equation reduces to 
\begin{align}
R_{\ \mu\sigma\nu}^{\rho}= & ^{c}R_{\ \mu\sigma\nu}^{\rho}+\nabla_{\sigma}D_{\ \mu\nu}^{\rho}-\nabla_{\nu}D^{\rho}{}_{\mu\sigma}\nonumber \\
 & +D^{\lambda}{}_{\mu\sigma}D^{\rho}{}_{\lambda\nu}-D^{\lambda}{}_{\mu\nu}D^{\rho}{}_{\lambda\sigma}+D^{\rho}{}_{\mu\lambda}\left(D^{\lambda}{}_{\nu\sigma}-D^{\lambda}{}_{\sigma\nu}\right).
\end{align}
\\

\subsection*{E.1 Metric teleparallel connection}

Metric teleparallel gravity \cite{AldrovandiPereira} is described via the following properties of the tensor fields: 
\begin{itemize}
\item $R^{\mu}{}_{\nu\rho\sigma}=0,$ 
\item $Q_{\mu\nu\rho}=0\mathbf{\ \ \ \ \ }\Rightarrow\mathbf{\ \ \ \ \ }L^{\mu}{}_{\rho\sigma}=0,$ 
\item $T^{\mu}{}_{\nu\sigma}\neq0\mathbf{\ \ \ \ \ }\Rightarrow\mathbf{\ \ \ \ \ }K^{\mu}{}_{\nu\sigma}\neq0.$ 
\end{itemize}
In this case, the contributions to the distortion tensor come
from the Levi Civita connection and the torsion tensor. Particularly,
one has the equality of the distortion and the contorsion. Namely
\begin{equation}
D^{\mu}{}_{v\sigma}=K^{\mu}{}_{v\sigma}.
\end{equation}
Since $R^{\mu}{}_{\nu\rho\sigma}=0$ and $D^{\mu}{}_{\nu\sigma}=K^{\mu}{}_{\nu\sigma}$
one has 
\begin{equation}
\begin{aligned}0=\  & ^{c}R^{\rho}{}_{\mu\sigma\nu}+\nabla_{\sigma}K^{\rho}{}_{\mu\nu}-\nabla_{\nu}K^{\rho}{}_{\mu\nu}+K^{\lambda}{}_{\mu\sigma}K^{\rho}{}_{\lambda\nu}\\
 & -K^{\lambda}{}_{\mu\nu}K^{\rho}{}_{\lambda\sigma}+K^{\rho}{}_{\mu\lambda}\left(K^{\lambda}{}_{\nu\sigma}-K^{\lambda}{}_{\sigma\nu}\right).
\end{aligned}
\end{equation}
Therefore, the Riemann tensor of the Levi-Civita connection, in terms of the contorsion tensor, reads 
\begin{align}
^{c}R^{\rho}{}_{\mu\sigma\nu} & =\nabla_{\nu}K_{\ \mu\sigma}^{\rho}-\nabla_{\sigma}K^{\rho}{}_{\mu\nu}+K^{\lambda}{}_{\mu\nu}K^{\rho}{}_{\lambda\sigma}\nonumber \\
 & -K^{\lambda}{}_{\mu\sigma}K^{\rho}{}_{\lambda\nu}+K^{\rho}{}_{\mu\lambda}\left(K^{\lambda}{}_{\sigma\nu}-K^{\lambda}{}_{\nu\sigma}\right).
\end{align}
By way of contraction, one obtains the Ricci tensor as 
\begin{align}
^{c}R_{\mu\nu}= & \nabla_{\nu}K^{\sigma}{}_{\mu\sigma}-\nabla_{\sigma}K^{\sigma}{}_{\mu\nu}+K^{\lambda}{}_{\mu\nu}K^{\sigma}{}_{\lambda\sigma}\\
 & -K^{\lambda}{}_{\mu\sigma}K^{\sigma}{}_{\lambda\nu}+K^{\sigma}{}_{\mu\lambda}\left(K^{\lambda}{}_{\sigma\nu}-K^{\lambda}{}_{\nu\sigma}\right).\nonumber 
\end{align}
The scalar curvature then becomes 
\begin{equation}
\begin{aligned}^{c}R= & \nabla_{\mu}K^{\sigma\mu}{}_{\sigma}-\nabla_{\sigma}K^{\sigma}{}_{\mu}{}^{\mu}+K^{\lambda}{}_{\mu}{}^{\mu}K^{\sigma}{}_{\lambda\sigma}\\
 & -K^{\lambda}{}_{\mu\sigma}K^{\sigma}{}_{\lambda}{}^{\mu}+K^{\sigma}{}_{\mu\lambda}\left(K^{\lambda}{}_{\sigma}{}^{\mu}-K^{\lambda\mu}{}_{\sigma}\right).
\end{aligned}
\end{equation}
Remember the antisymmetry property of the K-tensor: $K_{\mu\sigma\nu}=-K_{\sigma\mu\nu}$.
So, we can write 
\begin{equation}
\begin{gathered}^{c}R=2\nabla_{\mu}K^{\sigma\mu}{}_{\sigma}-K_{\mu}{}^{\lambda\mu}K^{\sigma}{}_{\lambda\sigma}-K_{\lambda\mu\sigma}K^{\sigma\lambda\mu}+K^{\sigma\mu\lambda}K_{\lambda\sigma\mu}-K^{\sigma\mu\lambda}K_{\lambda\mu\sigma},\end{gathered}
\end{equation}
or in a more compact form 
\begin{equation}
\begin{gathered}^{c}R=2\nabla_{\mu}K^{\sigma\mu}{}_{\sigma-}K_{\mu}{}^{\lambda\mu}K^{\sigma}{}_{\lambda\sigma}+K^{\sigma\mu\lambda}K_{\lambda\sigma\mu}-K_{\lambda\mu\sigma}\left(K^{\sigma\lambda\mu}+K^{\sigma\mu\lambda}\right),\end{gathered}
\end{equation}
where the last two terms automatically vanish because of the symmetries.
Then 
\begin{equation}
^{c}R=2\nabla_{\mu}K^{\sigma\mu}{}_{\sigma}-K_{\mu}{}^{\lambda\mu}K^{\sigma}{}_{\lambda\sigma}+K^{\sigma\mu\lambda}K_{\lambda\sigma\mu}.
\end{equation}
Using the explicit form 
\begin{equation}
K^{\beta}{}_{\mu\sigma}=\frac{1}{2}\left(T^{\beta}{}_{\mu\sigma}-T_{\mu}{}^{\beta}{}_{\sigma}-T_{\sigma}{}^{\beta}{}_{\mu}\right),
\end{equation}
one has $T^{\mu}{}_{\nu\sigma}=-T^{\mu}{}_{\sigma\nu}$, and so 
\begin{equation}
K_{\ \mu\sigma}^{\sigma}=\frac{1}{2}\left(T_{\ \mu\sigma}^{\sigma}-T_{\mu}^{\ \sigma}{}_{\sigma}-T_{\sigma}^{\ \sigma}{}_{\mu}\right)=\frac{1}{2}\left(T_{\ \mu\sigma}^{\sigma}-T_{\sigma}{}^{\sigma}{}_{\mu}\right),
\end{equation}
which reduces to 
\begin{equation}
K_{\ \mu\sigma}^{\sigma}=T_{\ \mu\sigma}^{\sigma}.
\end{equation}
Also, one has the following contraction: 
\begin{equation}
\begin{aligned}K^{\sigma\mu\lambda}K_{\lambda\sigma\mu}=\frac{1}{4} & \left(T^{\sigma\mu\lambda}-T^{\mu\sigma\lambda}-T^{\lambda\sigma\mu}\right)\left(T_{\lambda\sigma\mu}-T_{\sigma\lambda\mu}-T_{\mu\lambda\sigma}\right)\\
=\frac{1}{4} & \biggl(T^{\sigma\mu\lambda}T_{\lambda\sigma\mu}-T^{\sigma\mu\lambda}T_{\sigma\lambda\mu}-T^{\sigma\mu\lambda}T_{\mu\lambda\sigma}\\
 & -T^{\mu\sigma\lambda}T_{\lambda\sigma\mu}+T^{\mu\sigma\lambda}T_{\sigma\lambda\mu}+T^{\mu\sigma\lambda}T_{\mu\lambda\sigma}\\
 & -T^{\lambda\sigma\mu}T_{\lambda\sigma\mu}+T^{\lambda\sigma\mu}\left(T_{\sigma\lambda\mu}+T_{\mu\lambda\sigma}\right)\biggr),
\end{aligned}
\end{equation}
where the last two terms vanish because of the symmetries. Then we
have 
\begin{equation}
\begin{aligned}K^{\sigma\mu\lambda}K_{\lambda\sigma\mu}=\frac{1}{4} & \biggl(T_{\lambda\sigma\mu}(T^{\sigma\mu\lambda}-T^{\mu\sigma\lambda}-T^{\lambda\sigma\mu})\\
 & +T_{\sigma\lambda\mu}(T^{\mu\sigma\lambda}-T^{\sigma\mu\lambda})+T_{\mu\lambda\sigma}(T^{\mu\sigma\lambda}-T^{\sigma\mu\lambda})\biggr).
\end{aligned}
\end{equation}
By renaming the indices, it has become 
\begin{align}
K^{\sigma\mu\lambda}K_{\lambda\sigma\mu} & =\frac{1}{4}T_{\lambda\sigma\mu}\left(T^{\sigma\mu\lambda}-T^{\mu\sigma\lambda}-T^{\lambda\sigma\mu}+T^{\mu\lambda\sigma}-T^{\lambda\mu\sigma}+T^{\lambda\mu\sigma}-T^{\mu\lambda\sigma}\right),
\end{align}
which reduces to 
\begin{equation}
K^{\sigma\mu\lambda}K_{\lambda\sigma\mu}=\frac{1}{4}T_{\lambda\sigma\mu}\left(T^{\sigma\mu\lambda}-T^{\mu\sigma\lambda}-T^{\lambda\sigma\mu}\right).
\end{equation}
Using the antisymmetry of the torsion, one has 
\begin{equation}
K^{\sigma\mu\lambda}K_{\lambda\sigma\mu}=\frac{1}{4}\left(T_{\lambda\sigma\mu}T^{\sigma\mu\lambda}+T_{\lambda\mu\sigma}T^{\mu\sigma\lambda}-T_{\lambda\sigma\mu}T^{\lambda\sigma\mu}\right).
\end{equation}
Finally, one ends up with the following expression 
\begin{align}
K^{\sigma\mu\lambda}K_{\lambda\sigma\mu} & =\frac{1}{4}T_{\lambda\sigma\mu}\left(T^{\sigma\mu\lambda}-T^{\mu\sigma\lambda}-T^{\lambda\sigma\mu}\right)\nonumber \\
 & =\frac{1}{4}\left(2T^{\sigma\mu\lambda}-T^{\lambda\sigma\mu}\right)T_{\lambda\sigma\mu}.
\end{align}
Then, the scalar curvature can be rewritten as

\noindent\fbox{\begin{minipage}[t]{1\columnwidth - 2\fboxsep - 2\fboxrule}%
\begin{equation}
^{c}R=2\nabla_{\mu}K^{\sigma\mu}{}_{\sigma}-K_{\mu}{}^{\lambda\mu}K^{\sigma}{}_{\lambda\sigma}+K^{\sigma\mu\lambda}K_{\lambda\sigma\mu}=2\nabla_{\mu}T^{\sigma\mu}{}_{\sigma}-T_{\mu}{}^{\lambda\mu}T^{\sigma}{}_{\lambda\sigma}+\frac{1}{4}T_{\lambda\sigma\mu}\left(2T^{\sigma\mu\lambda}-T^{\lambda\sigma\mu}\right).
\end{equation}
\end{minipage}}

\noindent\\
 \\
 The last equation implies that the Einstein-Hilbert action 
\begin{equation}
S_{EH}=\frac{1}{2R}\int d^{4}x\sqrt{-g}{}^{c}R,
\end{equation}
can be rewritten in terms of the torsion tensor instead of the Christoffel
connection.\\

\subsection*{E.2 Symmetric teleparallel connection}

Symmetric teleparallel gravity \cite{BeltranJimenez} connection-satisfies the following relations: 
\begin{itemize}
\item $R^{\mu}{}_{\nu\rho\sigma}=0,$ 
\item $T^{\mu}{}_{\nu\rho}=0\mathbf{\ \ \ \ \ }\Rightarrow\mathbf{\ \ \ \ \ }K^{\mu}{}_{\nu\rho}=0,$ 
\item $Q_{\mu\nu\rho}\neq0\mathbf{\ \ \ \ \ }\Rightarrow\mathbf{\ \ \ \ \ }L^{\mu}{}_{\nu\rho}\neq0.$ 
\end{itemize}
In this case, the distortion is equal to the L-tensor, $D^{\mu}{}_{\nu\sigma}=L^{\mu}{}_{\nu\sigma}$.
Therefore, the Riemann tensor of the Christoffel connection can be
expressed as 
\begin{align}
^{c}R^{\rho}{}_{\mu\sigma\nu}= & \nabla_{\nu}L^{\rho}{}_{\mu\sigma}-\nabla_{\sigma}L^{\rho}{}_{\mu\nu}+L^{\lambda}{}_{\mu\nu}L^{\rho}{}_{\lambda\sigma}\\
 & -L^{\lambda}{}_{\mu\sigma}L^{\rho}{}_{\lambda\nu}+L^{\rho}{}_{\mu\lambda}\left(L^{\lambda}{}_{\sigma\nu}-L^{\lambda}{}_{\nu\sigma}\right),\nonumber 
\end{align}
and it yields the Ricci tensor of the Levi Civita connection as 
\begin{align}
^{c}R_{\mu\nu}= & \nabla_{\nu}L^{\sigma}{}_{\mu\sigma}-\nabla_{\sigma}L^{\sigma}{}_{\mu\nu}+L^{\lambda}{}_{\mu\nu}L^{\sigma}{}_{\lambda\sigma}\\
 & -L^{\lambda}{}_{\mu\sigma}L^{\sigma}{}_{\lambda\nu}+L^{\sigma}{}_{\mu\lambda}\left(L^{\lambda}{}_{\sigma\nu}-L^{\lambda}{}_{\nu\sigma}\right).\nonumber 
\end{align}
Note that we should be careful in raising the indices because of the
non-vanishing of the nonmetricity tensor. The connection is not metric
compatible, which means 
\begin{equation}
g^{\mu\nu}\nabla_{\sigma}L_{~\mu\nu}^{\sigma}\neq\nabla_{\sigma}L^{\sigma\mu}{}_{\mu}.
\end{equation}
The Ricci scalar becomes 
\begin{equation}
\begin{aligned}^{c}R=\nabla^{\mu} & L^{\sigma}{}_{\mu\sigma}-g{}^{\mu\nu}\nabla_{\sigma}L^{\sigma}{}_{\mu\nu}+L^{\lambda\mu}{}_{\mu}L^{\sigma}{}_{\lambda\sigma}\\
- & L^{\lambda}{}_{\mu\sigma}L^{\sigma}{}_{\lambda}{}^{\mu}+L^{\sigma}{}_{\mu\lambda}\left(L^{\lambda}{}_{\sigma}{}^{\mu}-L^{\lambda\mu}{}_{\sigma}\right).
\end{aligned}
\end{equation}
Remember that $L_{\mu\nu\sigma}$ is symmetric in its last two indices.
Hence, we can write 
\begin{equation}
\begin{aligned}^{c}R=\nabla^{\mu} & L_{\mu\sigma-}^{\sigma}g^{\mu\nu}\nabla_{\sigma}L_{\mu\nu}^{\sigma}+L^{\lambda\mu}{}_{\mu}L^{\sigma}{}_{\lambda\sigma}\\
 & -L_{\lambda\mu\sigma}L^{\sigma\lambda\mu}+L_{\sigma\mu\lambda}\left(L^{\lambda\sigma\mu}-L^{\lambda\sigma\mu}\right),
\end{aligned}
\end{equation}
which reduces to the following

\begin{equation}
^{c}R=\nabla^{\mu}L^{\sigma}{}_{\mu\sigma}-g^{\mu\nu}\nabla_{\sigma}L^{\sigma}{}_{\mu\nu}+L^{\lambda\mu}{}_{\mu}L^{\sigma}{}_{\sigma\lambda}-L_{\lambda\mu\sigma}L^{\sigma\lambda\mu}.
\end{equation}
 Using this result, one can express the scalar curvature of the Levi-Civita
connection in terms of $L$, and thus the Einstein-Hilbert action. That
formulation provides an alternative approach to general relativity.

\section*{APPENDIX F: TETRADS, THE SPIN CONNECTION, AND THE TELEPARALLEL GEOMETRY}

This appendix provides the minimal tetrad and the spin--connection
\cite{HehlReview} framework needed to interpret teleparallel 
gravity \cite{AldrovandiPereira} and to connect the metric--affine
wave equations derived in the main text with the gauge-theoretic
formulations of the spacetime geometry. No new dynamics are introduced here; the purpose is conceptual unification.

\subsection*{F.1 Tetrads and the spacetime metric}

We introduce the tetrad (vierbein) $e^{A}{}_{\mu}$ and its inverse
$E_{A}{}^{\mu}$, satisfying 
\begin{equation}
E_{A}{}^{\mu}e^{A}{}_{\nu}=\delta^{\mu}{}_{\nu},\qquad E_{A}{}^{\mu}e^{B}{}_{\mu}=\delta_{A}^{B}.
\end{equation}
The spacetime metric is expressed as 
\begin{equation}
g_{\mu\nu}=\eta_{AB}\,e^{A}{}_{\mu}e^{B}{}_{\nu},
\end{equation}
where $\eta_{AB}=\mathrm{diag}(1,-1,-1,-1)$ is the Minkowski metric.
This establishes the equivalence between the metric and the tetrad
descriptions of spacetime geometry.

\subsection*{F.2 Spin connection and the affine connection}

The spin connection \cite{HehlReview}, $\omega^{A}{}_{B\mu}$, relates
to the spacetime affine connection via 
\begin{equation}
\Gamma^{\rho}{}_{\mu\nu}=E_{A}{}^{\rho}\left(\partial_{\nu}e^{A}{}_{\mu}+\omega^{A}{}_{B\nu}e^{B}{}_{\mu}\right).
\end{equation}
This follows from the tetrad postulate 
\begin{equation}
\nabla_{\nu}e^{A}{}_{\mu}=\partial_{\nu}e^{A}{}_{\mu}+\omega^{A}{}_{B\nu}e^{B}{}_{\mu}-\Gamma^{\rho}{}_{\mu\nu}e^{A}{}_{\rho}=0,
\end{equation}
which ensures compatibility between spacetime and Lorentz indices.

\subsection*{F.3 Torsion and the curvature in differential form language}

The torsion and curvature two-forms are defined as 
\begin{equation}
T^{A}=de^{A}+\omega^{A}{}_{B}\wedge e^{B},\qquad R^{A}{}_{B}=d\omega^{A}{}_{B}+\omega^{A}{}_{C}\wedge\omega^{C}{}_{B}.
\end{equation}
Their spacetime components are related to the usual tensors by 
\begin{equation}
T^{A}{}_{\mu\nu}=e^{A}{}_{\rho}T^{\rho}{}_{\mu\nu},\qquad R^{A}{}_{B\mu\nu}=e^{A}{}_{\rho}E_{B}{}^{\sigma}R^{\rho}{}_{\sigma\mu\nu}.
\end{equation}

\subsection*{F.4 Teleparallel geometry}

Teleparallel gravity \cite{AldrovandiPereira,BeltranJimenez} is characterized
by 
\begin{equation}
R^{A}{}_{B}=0,\qquad Q_{\mu\nu\rho}=0,\qquad T^{\mu}{}_{\nu\rho}\neq0.
\end{equation}
The vanishing curvature implies that the spin connection is a pure gauge,
\begin{equation}
\omega^{A}{}_{B\mu}=(\Lambda^{-1})^{A}{}_{C}\,\partial_{\mu}\Lambda^{C}{}_{B}.
\end{equation}
In the Weitzenb\"{o}ck gauge, one may choose 
\begin{equation}
\omega^{A}{}_{B\mu}=0,
\end{equation}
so that all gravitational information is encoded in the torsion tensor.
This formulation is dynamically equivalent to general relativity and
underlies the teleparallel limit discussed in Appendix~E.

\subsection*{F.5 Bianchi identities}

In differential-form notation, the Bianchi identities take the compact form
\begin{equation}
DT^{A}=R^{A}{}_{B}\wedge e^{B},\qquad DR^{A}{}_{B}=0.
\end{equation}
In the teleparallel case, where $R^{A}{}_{B}=0$, this reduces to
\begin{equation}
DT^{A}=0.
\end{equation}
These relations are the exterior calculus counterparts of the component
Bianchi identities derived in Appendix~D.

This appendix clarifies how the same spacetime geometry may be described
equivalently in terms of curvature (Riemannian geometry), torsion
(teleparallel geometry), or mixed curvature--torsion structures (metric--affine
geometry). The equivalence underlying the wave equations derived in the main text supports the central theme of this work: geometry itself
propagates.


\section*{APPENDIX G: HISTORICAL AND CONCEPTUAL DEVELOPMENT OF THE CURVATURE,
THE TORSION, AND THE TELEPARALLEL GEOMETRIES}

This appendix provides a historical and conceptual overview of the
geometric ideas underlying the main body of the paper. Our aim is
not merely historical completeness, but to clarify why the sequence of curvature $\rightarrow$ Ricci identities $\rightarrow$ torsion and
nonmetricity $\rightarrow$ tetrads and spin connections $\rightarrow$ teleparallel formulations is both natural and unavoidable.

\subsection*{H.1 Riemann, curvature, and the non-commutativity of derivatives}

The modern concept of curvature originates in the seminal work of
\textbf{Bernhard Riemann} (1854) \cite{Riemann}, who introduced the idea that geometry is encoded in how distances change infinitesimally.
Riemann already understood that curvature measures the obstruction
to Euclidean behavior, but it was only later that this idea was given
its algebraic formulation.

In contemporary language, curvature is most cleanly characterized
by the non-commutativity of covariant derivatives: 
\begin{equation}
[\nabla_{\mu},\nabla_{\nu}]V^{\rho}=R^{\rho}{}_{\sigma\mu\nu}V^{\sigma}.
\end{equation}
This viewpoint reveals curvature not merely as a tensor, but as a
statement about the failure of parallel transport to be path-independent.
The Jacobi identity for differential operators then guarantees the
existence of nontrivial identities satisfied by the curvature tensor,
foreshadowing the Bianchi identities.

\subsection*{H.2 Ricci, Levi-Civita, and the classical tensor calculus}

The systematic development of tensor calculus was carried out by \textbf{Gregorio
Ricci-Curbastro} and \textbf{Tullio Levi-Civita} \cite{RicciLeviCivita}
at the turn of the twentieth century. Their construction culminated
in the introduction of the Levi-Civita connection, uniquely characterized
by 
\begin{equation}
\nabla_{\lambda}g_{\mu\nu}=0,\qquad T^{\rho}{}_{\mu\nu}=0.
\end{equation}
Within this framework, the Riemann tensor acquires additional symmetries,
and the Ricci identity and Bianchi identities take their familiar
forms. These identities are not optional mathematical decorations:
they arise directly from the algebraic structure of covariant differentiation
and ultimately encode the geometry's internal consistency.

\subsection*{H.3 Einstein and the dynamical role of the Bianchi identities}

In 1915, \textbf{Albert Einstein} elevated Riemannian geometry
from a mathematical structure to a physical theory of gravitation
\cite{Einstein1915}. By choosing the Levi-Civita connection \cite{RicciLeviCivita},
Einstein ensured that the contracted second Bianchi identity implies
\begin{equation}
\nabla_{\mu}G^{\mu\nu}=0,
\end{equation}
which, in turn, enforces the covariant conservation of the energy-momentum
tensor. Historically, this was a decisive insight: the Bianchi identities
were recognized as the geometric reason why Einstein's field equations
are dynamically consistent. Thus, the curvature identities became
the physical conservation laws.

\subsection*{H.4 Cartan geometry: torsion as an independent geometric field}

Shortly after the formulation of general relativity, \textbf{Elie
Cartan} \cite{Cartan} introduced a far-reaching generalization of
geometry. In Cartan's framework \cite{Cartan}, the affine connection
is no longer required to be symmetric or metric-compatible. Two new
geometric objects emerge: 
\begin{equation}
T^{\rho}{}_{\mu\nu}=2\Gamma^{\rho}{}_{[\mu\nu]},\qquad Q_{\lambda\mu\nu}=\nabla_{\lambda}g_{\mu\nu}.
\end{equation}
Cartan reformulated geometry in terms of differential forms, introducing
the coframe (tetrad) $e^{A}$ and the spin connection $\omega^{A}{}_{B}$.
Curvature and torsion appear naturally as field strengths, 
\begin{equation}
T^{A}=De^{A},\qquad R^{A}{}_{B}=d\omega^{A}{}_{B}+\omega^{A}{}_{C}\wedge\omega^{C}{}_{B},
\end{equation}
and satisfy the generalized Bianchi identities. This formalism unifies
the curvature, the torsion, and the nonmetricity into a single geometric
language.

\subsection*{H.5 Spinors, tetrads, and the necessity of the spin connection}

The introduction of spinor fields by \textbf{Dirac} in 1928 \cite{Dirac}
revealed a crucial limitation of purely metric formulations of gravity.
Spinors transform under local Lorentz transformations rather than
general coordinate transformations, necessitating the introduction
of tetrads and spin connections even within general relativity.

Thus, the tetrad--spin connection formalism is not an alternative
description of gravity, but the natural framework once fermionic matter
is included. Appendix F establishes this structure and shows how the
spacetime tensors and the Lorentz-indexed objects are consistently
related.

\subsection*{H.6 Teleparallel gravity and Einstein's alternative viewpoint}

In an attempt to unify gravity with electromagnetism, Einstein explored
teleparallel or distant parallelism geometries in the late 1920s \cite{AldrovandiPereira}.
Although the unification program was abandoned, a key insight survived:
the gravitational interaction can be encoded entirely in torsion,
with vanishing curvature.

The modern teleparallel equivalent of general relativity (TEGR) is characterized
by 
\begin{equation}
R^{\rho}{}_{\sigma\mu\nu}=0,\qquad Q_{\lambda\mu\nu}=0,\qquad T^{\rho}{}_{\mu\nu}\neq0.
\end{equation}
Appendix~E explicitly demonstrates how the Einstein--Hilbert scalar
curvature can be rewritten as a torsion-squared invariant plus a boundary
term. Thus, curvature and torsion provide equivalent descriptions
of the same gravitational dynamics.

\subsection*{H.7 Symmetric teleparallel gravity and nonmetricity}

A complementary formulation, known as symmetric teleparallel gravity,
sets both curvature and torsion to zero while retaining nonmetricity:
\begin{equation}
R^{\rho}{}_{\sigma\mu\nu}=0,\qquad T^{\rho}{}_{\mu\nu}=0,\qquad Q_{\lambda\mu\nu}\neq0.
\end{equation}
In this case, gravity is entirely encoded in the failure of the connection
to preserve the metric. This formulation completes a geometric triad: 
\begin{center}
\begin{tabular}{ccc}
Curvature  & Torsion  & Nonmetricity \tabularnewline
$\neq0$  & $=0$  & $=0$ \tabularnewline
$=0$  & $\neq0$  & $=0$ \tabularnewline
$=0$  & $=0$  & $\neq0$ \tabularnewline
\end{tabular}
\par\end{center}

The historical development reviewed here reveals a unifying theme: 
\begin{itemize}
\item Gravity can be understood as geometry, but the geometry itself admits
multiple equivalent encodings. 
\item Curvature, torsion, and nonmetricity represent different ways in which
covariant derivatives fail to commute or preserve structure. The appendices
of this paper make explicit how these descriptions are related and
how general relativity emerges as a special, but not unique, geometric
choice. For recent studies in various geometries, see \cite{Adak1,Adak2,Bahamonde,Sotiriou,Heisenberg2,Jarv,Barrientos}
and the references therein. 
\end{itemize}
\begin{acknowledgments}
We thank Muzaffer Adak for useful discussions and for a careful reading of the manuscript. We are also grateful to Friedrich W.~Hehl for an important remark that led to a revision of the table.

Professor Metin G\"{u}rses has been a constant source of inspiration through his never-fading enthusiasm for physics, his exceptional work ethic, and his generosity as a collaborator and colleague. Having had the privilege of working with him for more than a decade on research and many joint papers, we deeply appreciate both his scientific insight and his kindness as a person. On the occasion of his becoming an emeritus professor at the Bilkent University Department of Mathematics, we celebrate his remarkable contributions to gravitational and mathematical physics and wish him many more years of joyful and fruitful exploration of gravity.
  
E.~Altas is supported by the TUBITAK Grant No.~123F353.
\end{acknowledgments}

\end{document}